\documentclass[trackchanges]{aastex701}

\usepackage{amsmath}

\begin{document}

\title{Plasmoid-Trapped Condensation Associated with a Transient Thermal-Instability-Like Process in Chromospheric Magnetic Reconnection}

\author[orcid=0000-0002-2388-7068,sname='Abdullah Zafar']{Abdullah Zafar}
\affiliation{Yunnan Observatories, Chinese Academy of Science, Kunming, Yunnan 650216, People’s Republic of China}
\email[show]{abdallahzafar@gmail.com}  

\author[orcid=0000-0001-6366-7724, sname='Lei Ni']{Lei Ni} 
\affiliation{Yunnan Observatories, Chinese Academy of Science, Kunming, Yunnan 650216, People’s Republic of China}
\affiliation{Yunnan Key Laboratory of Solar Physics and Space Science, Kunming 650216, People’s Republic of China}
\affiliation{University of Chinese Academy of Sciences, Beijing 100049, People’s Republic of China}
\email[show]{leini@ynao.ac.cn}

\author[orcid=0000-0003-1546-381X,sname='Samrat Sen']{Samrat Sen}
\affiliation{Dr. K. S. Krishnan Geomagnetic Research Laboratory, Indian Institute of Geomagnetism, Prayagraj, India}
\email[]{}  

\author[orcid=0000-0002-8077-094X,sname='Yuhao Chen']{Yuhao Chen}
\affiliation{School of Earth and Space Sciences, Peking University, Beijing 100781, People’s Republic of China}
\affiliation{State Key Laboratory of Solar Activity and Space Weather, National Space Science Center, Chinese Academy of Sciences, Beijing
100190, People’s Republic of China}
\email[]{}  

\author[orcid=0009-0001-0996-8232,sname='Mohamed Sedik']{Mohamed Sedik}
\affiliation{Yunnan Observatories, Chinese Academy of Science, Kunming, Yunnan 650216, People’s Republic of China}
\email[]{}  

\author[orcid=0000-0003-2682-9784,sname='Mehdi Yousefzadeh']{Mehdi Yousefzadeh}
\affiliation{Yunnan Observatories, Chinese Academy of Science, Kunming, Yunnan 650216, People’s Republic of China}
\email[]{} 

\author[orcid=0000-0001-6366-7724, sname='Jun Lin']{Jun Lin} 
\affiliation{Yunnan Observatories, Chinese Academy of Science, Kunming, Yunnan 650216, People’s Republic of China}
\affiliation{Yunnan Key Laboratory of Solar Physics and Space Science, Kunming 650216, People’s Republic of China}
\affiliation{University of Chinese Academy of Sciences, Beijing 100049, People’s Republic of China}
\email{jlin@ynao.ac.cn}







\begin{abstract}

We perform high-resolution 2.5D single-fluid magnetohydrodynamic (MHD) simulations to explore the development of cool-dense structures within plasmoids during chromospheric magnetic reconnection. The model incorporates temperature-dependent ionization degrees of hydrogen and helium, resulting in improved diffusivities and viscosity, as well as adequate radiative cooling. The numerical
results show that plasma is significantly hot with temperatures reaching several tens of thousands of Kelvin in newly developed plasmoids in the low initial plasma-$\beta$ cases, $\beta_0$ = 0.05. Later, the localized accumulation of plasma and decreasing temperature result in an explosive much stronger radiative
cooling process. Radiative cooling peaks over the temperature range 8000 K$-$20,000 K. The co-spatial density enhancement, substantial radiative cooling, shorter radiative cooling timescales, and a significant temperature decrease support the existence of thermal-instability-like condensation within the
plasmoids. Similar cool-dense structures are identified in all low-$\beta_0$ cases at different chromospheric altitudes, suggesting that this occurrence is not limited to a single chromospheric layer. In contrast, in
the high initial plasma-$\beta$, $\beta_0$ = 0.5 case, the maximum temperature inside the dense plasmoids is only about 8000 K. Therefore, the radiative cooling has never entered into the explosive increasing stage and the thermal-instability-like condensation does not happen. These findings reveal that plasmoid-trapped
condensation may develop in chromospheric reconnection once radiative losses become self-amplifying under low-$\beta_0$ conditions.

\end{abstract}



\section{Introduction} 
Magnetic reconnection is a ubiquitous process in magnetized plasmas that powers many energetic and explosive
events in the Universe. These include coronal mass ejections and solar flares in our solar system~\citep{parker1963solar,masuda1994loop,yamada2010magnetic}, as well as similar high-energy reconnection events in the coronae
of other stars, young stellar objects~\citep{goodson1997time,feigelson1999high,benz2010physical}, and the interstellar medium~\citep{zweibel1989magnetic,brandenburg1995effects,heitsch2003fast}.
Reconnection occurs when plasma flows advect anti-parallel magnetic field lines toward each other, resulting in the formation of a thin current sheet. 
The frozen-in flux condition breaks down locally within this thin layer, which alters the magnetic field topology and converts stored magnetic energy to thermal and kinetic energy of the plasma~\citep{yamada2010magnetic,ji2022magnetic}. 
Magnetic reconnection has been extensively investigated in fully ionized plasmas over the past several decades~\citep{zweibel2009magnetic,cassak2012magnetic}, however, its behavior in cooler and partially ionized environments remains less understood. 
The presence of neutral particles, which are a key constituent of partially ionized plasmas, introduces additional physical effects, such as their interaction with charged species, which influences current-sheet evolution, and is thus regarded as one of the ten major challenges in the magnetic reconnection process~\citep{ni2020magnetic}. 
The widespread presence of partially ionized plasmas in various astrophysical environments such as the lower solar atmosphere, comet tails, young stellar objects, protoplanetary nebulae, and the interstellar medium~\citep{ballester2018partially}, provides strong motivation to better explore the dynamics and evolution of such cool plasma settings.

The chromosphere, a thin layer between the photosphere and corona, is highly a dynamics and complex region that plays an important role in the mass and energy balance of the outer solar atmosphere~\citep{athay1982role,de2011origins}. 
Small-scale transient events, such as chromospheric jets~\citep{shibata2007chromospheric,singh2011chromospheric,cai2019multiband} and ultraviolet (UV) bursts~\citep{peter2014hot,tian2016iris,chen2019flame,Zhenghua2019observations}, are commonly observed in this region. Such transient and energetic events are believed to be triggered by magnetic reconnection. 
Localized outflows in the chromospheric jets are presumed to be the result of magnetic reconnection~\citep{shibata2007chromospheric}.
The identification of plasma bubbles in the outflow of UV bursts and chromospheric jets, which are usually regarded to be plasmoids, offers additional evidence of magnetic reconnection~\citep{singh2011chromospheric,guo2020observations}. 
A plasmoid, defined as a region of closed magnetic field lines that have an O-type null point located in the center of the 2D plane, develops when the Lundquist number ($S$) is larger than the critical value of $\sim$ 10$^4$~\citep{loureiro2007instability,bhattacharjee2009fast,ni2010linear}. 
They are frequently observed in various reconnecting solar settings~\citep{lin2005direct,takasao2012simultaneous,li2016magnetic,kumar2018evidence,kumar2019first,cheng2023ultra,cheng2024evidence,hou2024numerous,liu2026bidirectional,li2026unprecedented}. 
Plasmoid instability is essential in fast magnetic reconnection as it breaks a long current sheet into many plasmoids and secondary X-points, making the reconnection process highly dynamic~\citep{shibata2001plasmoid,loureiro2007instability,bhattacharjee2009fast}. 
Plasmoid-driven reconnection accelerates the reconnection process, while the associated turbulent flows and frequent plasmoid coalescence modify the temperature and density distributions, local heating, and plasma compression within the current sheet~\citep{ni2015fast,ni2016heating,ni2022plausibility,murtas2021coalescence,zafar2023high,zafar2025effect}.

The formation of localized dense and cool structures across various astrophysical environments, such as the solar atmosphere, interstellar clouds, planetary nebulae, and star-forming regions, has been widely attributed to a fundamental process in radiating astrophysical plasmas known as thermal instability~\citep{parker1953instability,field1965thermal,smith1977formation,forbes1991numerical,fang2015coronal,kohutova2020self,hennebelle2009diffuse,dudorov2019magnetic}. 
If a perturbation introduces a local thermal imbalance in a localized plasma, the resulting enhancement of radiative energy losses relative to the net heating rate can destabilize the system. Under isobaric, isochoric, or isentropic evolution, this imbalance may be further amplified, leading to progressively stronger cooling and ultimately triggering a thermally unstable runaway condensation process. 
Similar rapid cooling and condensation associated with thermal instability and thermal nonequilibrium have been widely investigated in the solar corona in relation to coronal rain, prominences, and post-flare loop formation~\citep{antiochos1999dynamic,antiochos2000thermal,karpen2005prominence,karpen2008condensation,xia2012simulations,fang2013multidimensional,keppens2014dynamics,li2022coronal,antolin2022thermal,keppens2025modeling,liakh2025numerical}.
Condensation formation in the vicinity of coronal X-point magnetic structures has also been investigated numerically~\citep{johnston2025filament}.
Recent 2D and 3D MHD simulations of reconnecting coronal current sheets showed that both tearing and thermal processes influence current sheet fragmentation, with tearing instability leading to the development of plasmoids or flux ropes, the neighboring plasmoid or flux ropes interact with each other, trap plasma, and provide favorable conditions for localized radiative condensation~\citep{sen2022thermally,sen20233d,de2025coupled}. 
In particular, these investigations found that condensations form during the nonlinear phase of tearing-dominated evolution, which provides a framework for exploring whether similar cool-dense structures associated with thermal-instability-like process may develop in plasmoid-dominated reconnection in the partially ionized chromosphere.

This work aims to explore the development and evolution of dense-cool structures caused by thermal-instability-like phenomena within hot plasmoids during plasmoid-mediated reconnection at various chromospheric altitudes. 
The combined effects of a more realistic radiative cooling model and temperature-dependent hydrogen and helium ionizations, which result in physically realistic magnetic diffusions, ambipolar diffusion, and viscosity, improve our understanding of the physical processes underlying thermal-instability-like phenomena in partially ionized chromospheric plasma.
This research will improve our understanding of the physical processes lead to dense-cool structures in partially ionized astrophysical environments. 
The structure of this paper is as follows. Section~\ref{sec_II} describes the simulation model and initial setup. The numerical results are discussed in Section~\ref{sec-III}. Finally, Section~\ref{sec-IV} provides a summary and conclusion of this work.

\section{Numerical model and initial setup}
\label{sec_II}
To understand the evolution of a reconnecting current sheet, we performed 2.5D MHD simulations using the single-fluid MHD code NIRVANA~\citep{ziegler2011semi}. All species of the partially ionized hydrogen-helium plasma (H, H$^{+}$, H$_{e}$, H$_{e}^{+}$ and electrons) are strongly coupled and thus considered as a single fluid~\citep{ni2022plausibility}. 
In this study, NIRVANA solves the following set of single-fluid MHD equations in Cartesian geometry. 
\begin{eqnarray}
\frac{\partial \rho}{\partial t} = - \nabla \cdot (\rho \mathbf{v}),
\label{eq:1}
\end{eqnarray}

\begin{eqnarray}
\begin{split}   
\frac{\partial (\rho \mathbf{v})}{\partial t} & = - \nabla \cdot \left[\rho \mathbf{vv}+\left(p+\frac{1}{2\mu_{0}} |\mathbf{B}|^{2}\right)I-\frac{1}{\mu_{0}}\mathbf{BB}\right] \\
&+\nabla \cdot \tau_{S},
\end{split}
\label{eq:2}
\end{eqnarray}

\begin{eqnarray}
\begin{split}
    \frac{\partial e}{\partial t} &= -\nabla \cdot \left[\left(e+p+\frac{1}{2\mu_{0}} |\mathbf{B}|^{2}\right) \mathbf{v}\right] \\
    & + \nabla \cdot \left[\frac{1}{\mu_{0}} (\mathbf{v} \cdot \mathbf{B})\mathbf{B}\right] \\
    & + \nabla \cdot \left[\mathbf{v} \cdot \tau_{s} + \frac{\eta}{\mu_{0}} \mathbf{B} \times (\nabla \times \mathbf{B})\right] \\
    & - \nabla \cdot \left[\frac{1}{\mu_{0}} \mathbf{B} \times \mathbf{E}_{AD}\right] \\
    & + Q_{rad} +\mathcal{H}, 
\end{split}
\label{eq:3}
\end{eqnarray}

\begin{eqnarray}
\frac{\partial \mathbf{B}}{\partial t} = \nabla \times (\mathbf{v} \times \mathbf{B} - \eta \nabla \times \mathbf{B} + \mathbf{E}_{AD}),
\label{eq:4}
\end{eqnarray}
with
\begin{eqnarray}
e = \frac{p}{\gamma -1} + \frac{1}{2} \rho |\mathbf{v}|^{2} + \frac{1}{2 \mu_{0}} |\mathbf{B}|^{2}
\label{eq:5}
\end{eqnarray}
and
\begin{eqnarray}
p = \frac{(1.1+Y_{iH}+0.1Y_{iHe})\rho}{1.4 m_{i}} k_{B} T,
\label{eq:6}
\end{eqnarray}
where $\rho$ is the mass density of plasma, $\mathbf{v}$ is the fluid velocity, $\mathbf{B}$ is the magnetic field, $p$ is the thermal pressure, $e$ denotes the total energy density, $T$ represents temperature; and $Y_{iH}$ and $Y_{iHe}$ are the temperature-dependent ionization fractions of hydrogen and helium, respectively, while m$_i$ is the proton mass, and $k_{B}$ is the Boltzmann constant. Helium’s total number density is 10\% that of hydrogen, and the first ionization of helium is considered only. The adiabatic constant ($\gamma$) is set to 5/3. The stress tensor, $\tau_{S} = \xi [\nabla \mathbf{v} + (\nabla \mathbf{v})^{T} - \frac{2}{3} (\nabla \cdot \mathbf{v})I]$ , where $\xi$ denotes the coefficient of dynamic viscosity, expressed in kg m$^{-1}$ s$^{-1}$. In this work, the current sheet is aligned parallel to the Sun’s surface, and its width narrows down to less than several tens of kilometers during the main reconnection process; thus, the gravity effect is not considered, and the initial plasma density and temperature are assumed to be uniform across the whole simulation domain.

The temperature dependent coefficients of magnetic diffusions due to electron-ion ($\eta_{ei}$) and electron-neutral ($\eta_{en}$)
collisions, ambipolar diffusion ($\eta_{AD}$), and dynamical viscosity ($\xi$) employed in this work are adopted from our previous
studies~\citep{zafar2023high,zafar2024unraveling,zafar2025effect} and are given as follows:
\begin{eqnarray}
\eta_{ei} \simeq 1.0246 \times 10^{8} \Lambda T^{-1.5},
\label{eq:eta_ei}
\end{eqnarray}
\begin{eqnarray}
\eta_{en} \simeq 0.0351 \sqrt{T} \frac{\left[\frac{0.1}{3} (1-Y_{iH_{e}})+(1-Y_{iH})\right]}{Y_{iH} + 0.1 Y_{iHe}},
\label{eq:eta_en}
\end{eqnarray}
\begin{eqnarray}
\eta_{AD} = \frac{(\rho_{n}/\rho)^2}{\rho_{i} \nu_{in}}.
\label{eq:eta_ad}
\end{eqnarray}
and 
\begin{eqnarray}
\xi =  \frac{4.8692 \times 10^{-16}}{\Lambda} T^{2} \sqrt{T}+ 2.0127 \times 10^{-7} \sqrt{T}. 
\label{eq:vis}
\end{eqnarray}
where $\rho_{n}$ and $\rho_{i}$ denote the neutral and ion mass densities, while $\Lambda$ and $\nu_{in}$ are the Coulomb logarithm and ion-neutral collision frequency, respectively.

Radiative transfer or the energy exchange between the radiation field and solar atmosphere, influences the thermodynamic behavior of plasma in the lower solar atmosphere. 
In order to model the numerical experiments of chromospheric magnetic reconnection processes more realistically, we utilize the Carlsson \& Leenaarts radiative cooling model~\citep{carlsson2012approximations}, a widely accepted approximation for radiative losses in the chromosphere and is given by
\begin{eqnarray}
Q_{rad} = - \sum_{X = H, Mg, Ca} L_{Xm} (T) E_{Xm} (\tau) \frac{N_{Xm}}{N_{X}} (T) A_{X} \frac{N_{H}}{\rho} n_{e} \rho,
\label{eq:rad}
\end{eqnarray}
where $L_{Xm} (T)$ represents the optically thin radiative loss function that changes with temperature, per electron, and per particle of element $X$ in the ionization phase $m$, 
$E_{Xm} (\tau)$ refers to the escape probability which is a function of
optical depth $\tau$,
$\frac{N_{Xm}}{N_{X}}(T)$ corresponds to the fraction of element $X$ in the ionization stage $m$, $A_X$ refers to the abundance of element $X$,
and $\frac{N_{H}}{\rho} = 4.407 \times 10^{23} g^{-1}$ gives the number of hydrogen particles per unit mass of the chromospheric material.
Following~\citet{carlsson2012approximations}, the total column density of the neutral hydrogen is multiplied by $4.0 \times 10^{-14}$ cm$^2$ to obtain a depth variable that is close to the optical depth at the Ly\(\alpha\) line center.
The quantities $L_{Xm}$, $E_{Xm}$, and $N_{Xm}/N_{X}$ were computed by \citet{carlsson2012approximations} from detailed radiative-transfer calculations using RADYN~\citep{1997ApJ...481..500C,2002ApJ...572..626C} for hydrogen, and BIFROST-based atmospheric models combined with MULTI3D radiative-transfer calculations~\citep{leenaarts2009multi3d} for Mg II and Ca II.
In the present simulations, the corresponding tabulated values provided by~\cite{carlsson2012approximations} are implemented in the NIRVANA code. 
At each grid cell and time step, $L_{Xm}$ and $N_{Xm}/N_{X}$ are obtained by interpolation from these tabulated data according to the local plasma temperature.
As aforementioned, the reconnecting current sheet gets extremely thin during the simulation, with a width of just a few tens of kilometers. We thus assume that the escape probability is fixed at each chromospheric height and remains unchanged during the simulation.
These simulations do not include any background heating (i.e., $H=0$). Q$_{rad}$ is also turned off when T $\leq$ T$_0$, where T$_0$ represents the initial temperature at each chromospheric height. Hence, both terms are initially zero.

2.5D simulations are performed with different initial plasma-$\beta$ values at various chromospheric altitudes, ranging from the bottom to the middle of the chromosphere, to analyze the formation of cool-dense structures associated with thermal instability-like processes in the reconnection current sheet. 
The numerical experiments are initialized using a magnetic field configuration corresponding to the force-free Harris current sheet, stated as $B_{x0} = -b_{0} \tanh [y/(0.05L_{0})]$, $B_{y0} = 0$, and $B_{z0} = b_{0}/ \cosh [y/(0.05L_{0})]$, where $b_0$ corresponds to the strength of the initial magnetic field.
At the center of the current sheet, the guide field reaches its maximum value, \(B_{Z,\max}=b_0\). The adopted force-free Harris-sheet equilibrium therefore represents an idealized strong-guide-field configuration rather than a typical chromospheric guide-field configuration; configurations with different guide-field ratios are beyond the scope of the present setup.
The initial plasma-$\beta$ changes with the strength of the initial magnetic field. 
The characteristic length \(L_0=2\times10^5\) m is chosen to represent the spatial scale of the local reconnection region considered in our simulations. A comparable value has been adopted in previous numerical studies of lower-atmospheric magnetic reconnection; for example,~\citet{liu2023numerical} used \(L_0=2\times10^5\) m for Ellerman-bomb reconnection and obtained radiative losses comparable to those inferred for a chromospheric reconnection event~\citep{diaz2021observationally}. 
High-resolution observations have likewise identified plasmoid-like structures associated with Ellerman bombs on sub-arcsecond spatial scales of about \(0.1''-0.4''\)~\citep{rouppe2023ultra}.
The simulation domain is from 0 to $L_0$ on the $x$-axis while ranging from $-0.5L_0$ to $0.5L_0$ along the $y$-axis.  
Inflow boundary conditions are applied along the y-direction, while outflow boundary conditions are imposed along the x-direction, allowing plasmoids to leave the computational domain after reaching the ends of the current sheet~\citep{ni2021magnetic}.
To trigger magnetic reconnection, a small initial magnetic perturbation is introduced, expressed as $b_{x1} = - b_{pert} \sin \left[{2 \pi (y+0.5L_{0})}/{L_{0}}\right] \cos \left[{2 \pi (x+0.5L_{0})}/{L_{0}}\right]$ and $b_{y1} = b_{pert} \cos \left[{2 \pi (y+0.5L_{0})}/{L_{0}}\right] \sin \left[{2 \pi (x+0.5L_{0})}/{L_{0}}\right]$ with b$_{pert}$  = 0.005 b$_{0}$. 
In this work, a base-level grid of 192 × 192 is employed with an adaptive mesh refinement (AMR) level of 9, resulting in a minimum grid size of approximately 2 m.

\section{Numerical results}
\label{sec-III}

\begin{figure}
\centering
\begin{minipage}{0.494\textwidth}
\includegraphics[width=1.0\textwidth]{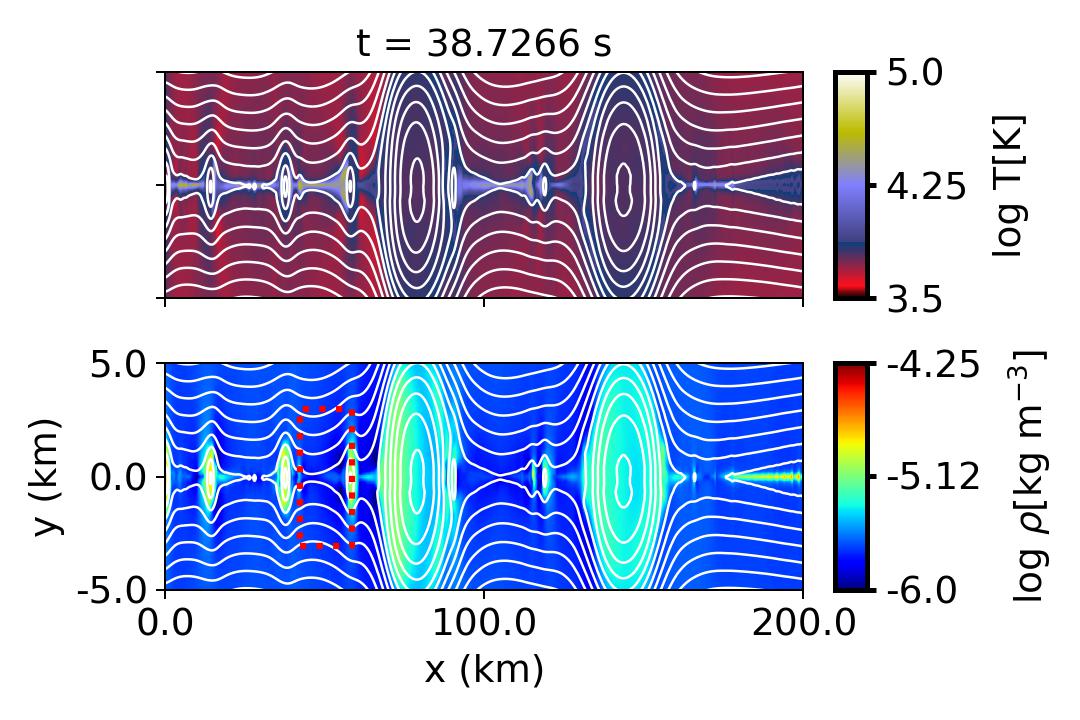}
\put(-80,142){\textcolor{white}{\textbf{(a)}}}
\end{minipage}
\begin{minipage}{0.494\textwidth}
\includegraphics[width=1.0\textwidth]{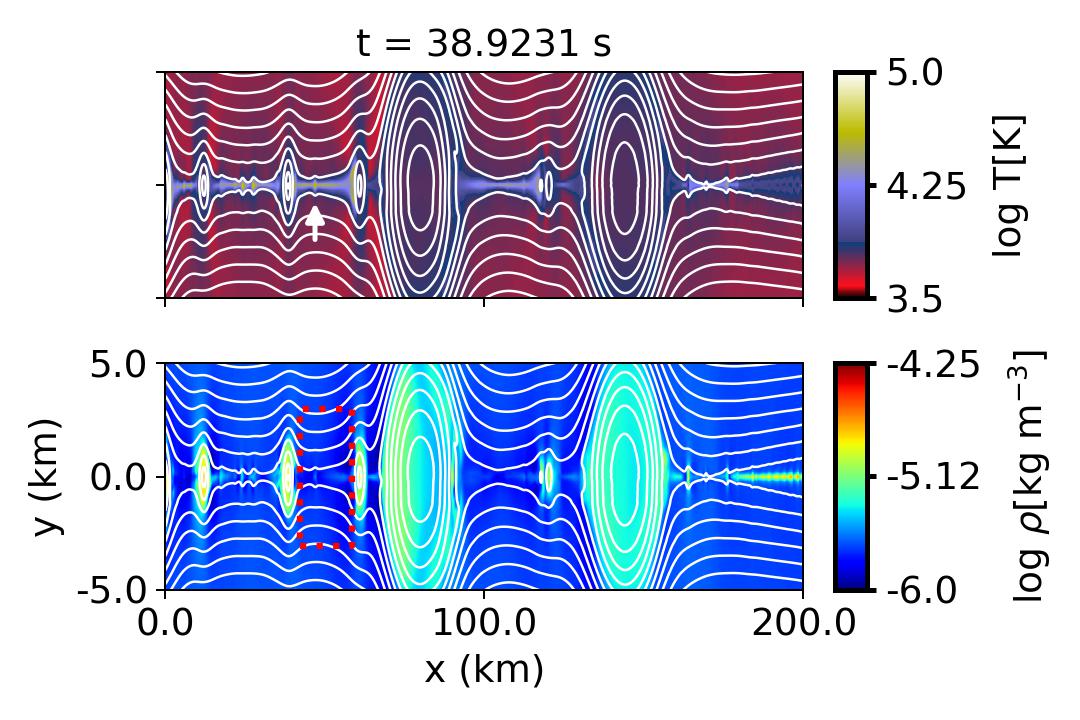}
\put(-80,142){\textcolor{white}{\textbf{(b)}}}
\end{minipage}
\caption{Global view of the distribution of temperature in the top panels and density in the bottom panels at two different times during magnetic reconnection at the bottom of the chromosphere ($Z = 600$ km) with initial plasma-$\beta$ of 0.05. White lines indicate magnetic field lines. The red box in the density panels outlines the region of interest analyzed in detail in Figures~\ref{fig_2} and \ref{fig_3}. The white arrow in the right temperature panel marks the location of the newly born plasmoid, later denoted as P1.}
\label{fig_1}
\end{figure}

Figure~\ref{fig_1} presents a global view of the nonlinear stage of the reconnecting current sheet at the base of the chromosphere, 600 km above the solar surface with an initial plasma-$\beta$ $\sim$ 0.05 ($b_0 \sim 5.9 \times 10^{-2}$ T). 
At this altitude, the initial plasma mass density ($\rho_0$ ) and temperature (T$_0$ ) are $2.4 \times 10^{-6}$ kg m$^{-3}$ and 4421 K, respectively. 
These parameters ($\rho_0$ , T$_0$) are determined from the C7 atmosphere model~\citep{avrett2008models}. 
The contours show density (upper panel) and temperature (lower panel) distributions at two different times in the full horizontal domain, and
the vertical domain is displayed in the range $-5$ km $\leq y \leq$ 5 km to clearly identify the structures emerging in the reconnection region. 
The white lines represent the magnetic field lines. Similar to a typical chromospheric reconnection scenario, we can see a chain of small and big plasmoids, and both temperature and density are highly non-uniform along the current sheet. 
Figure~\ref{fig_1}(a) demonstrates that there is no plasmoid or localized plasma structure within the thin current sheet enclosed by the red dotted box, despite the surrounding current sheet being highly structured outside and around this area of interest. 
However, by $t = 38.92$ s, a plasmoid forms within the region enclosed by the red dotted box, as represented by the white arrow in the panel of Figure~\ref{fig_1}(b). 
This structure is evident in the temperature and density panels, where the newly developed plasmoid exhibits both an elevated temperature and a localized density enhancement. 
At this early stage, the high temperature of the plasmoid is most likely due to the accumulation of hot plasma outflow from the surrounding X-points, as the structure has not evolved enough for radiative cooling to be effective. 
These two frames identify the emergence of a plasmoid within the area of interest (red dotted box), providing a framework for the subsequent investigation.

\begin{figure}
\centering
\begin{minipage}{0.3\textwidth}
\includegraphics[width=1.0\textwidth]{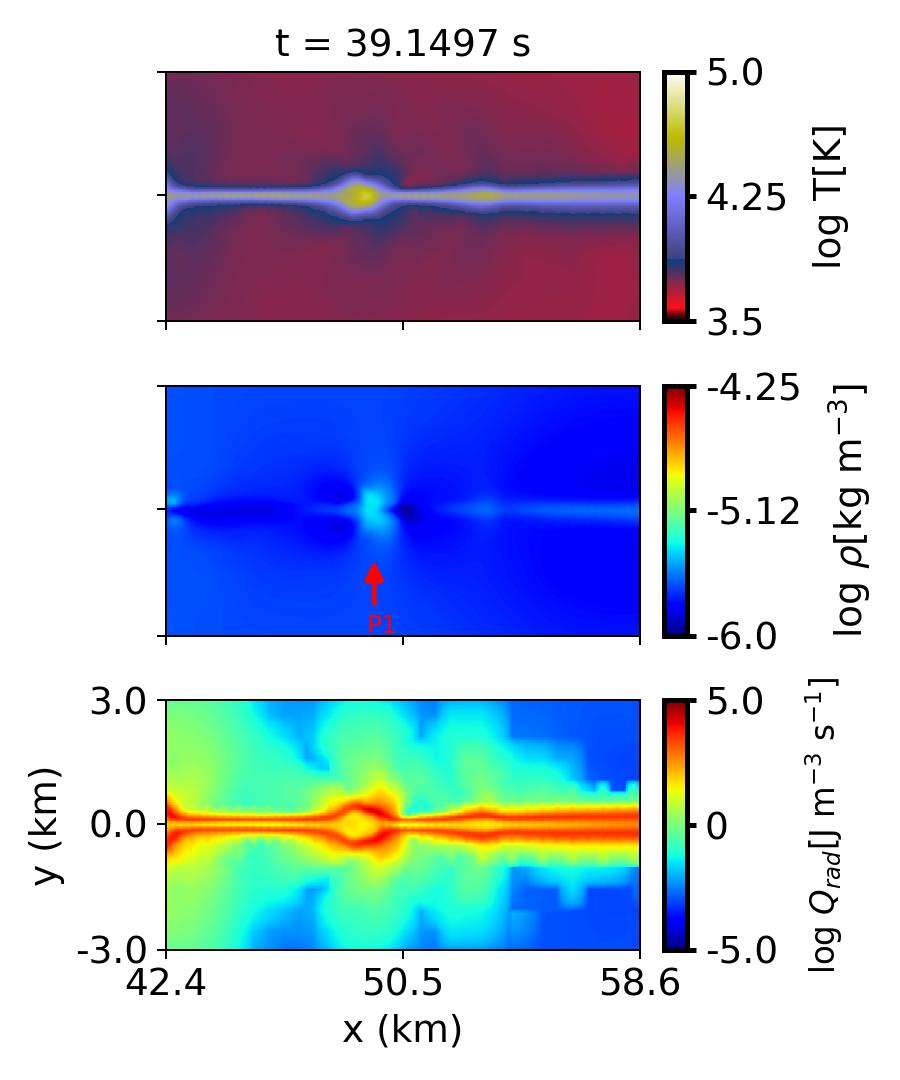}
\put(-60,165){\textcolor{white}{\textbf{(a)}}}
\end{minipage}
\begin{minipage}{0.3\textwidth}
\includegraphics[width=1.0\textwidth]{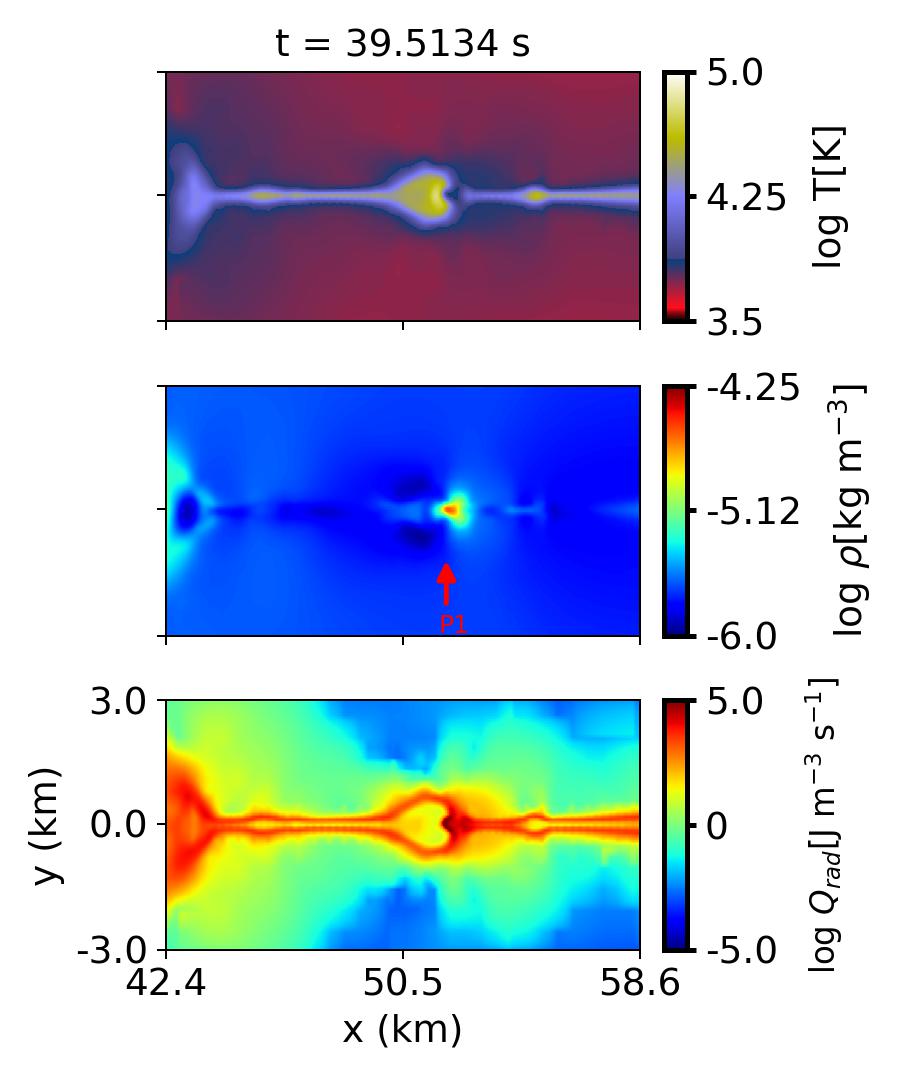}
\put(-60,165){\textcolor{white}{\textbf{(b)}}}
\end{minipage}
\begin{minipage}{0.3\textwidth}
\includegraphics[width=1.0\textwidth]{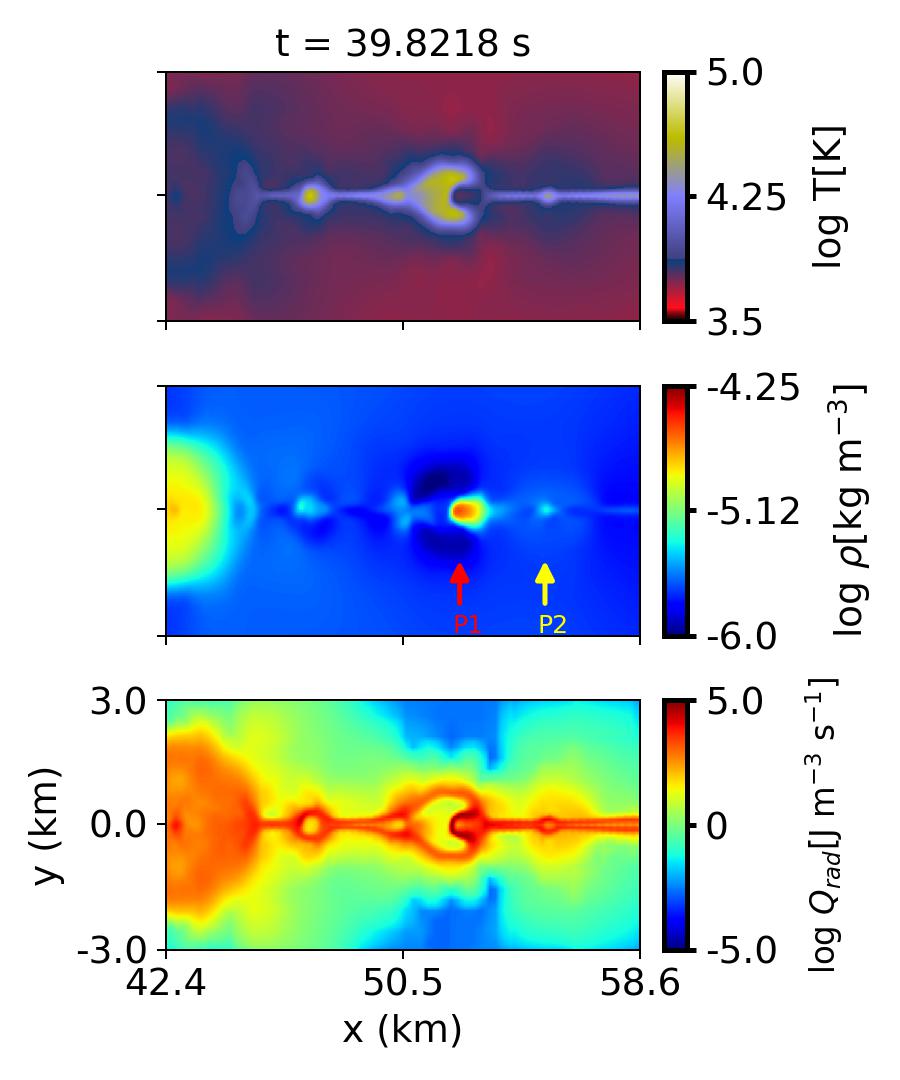}
\put(-60,165){\textcolor{white}{\textbf{(c)}}}
\end{minipage}
\begin{minipage}{0.3\textwidth}
\includegraphics[width=1.0\textwidth]{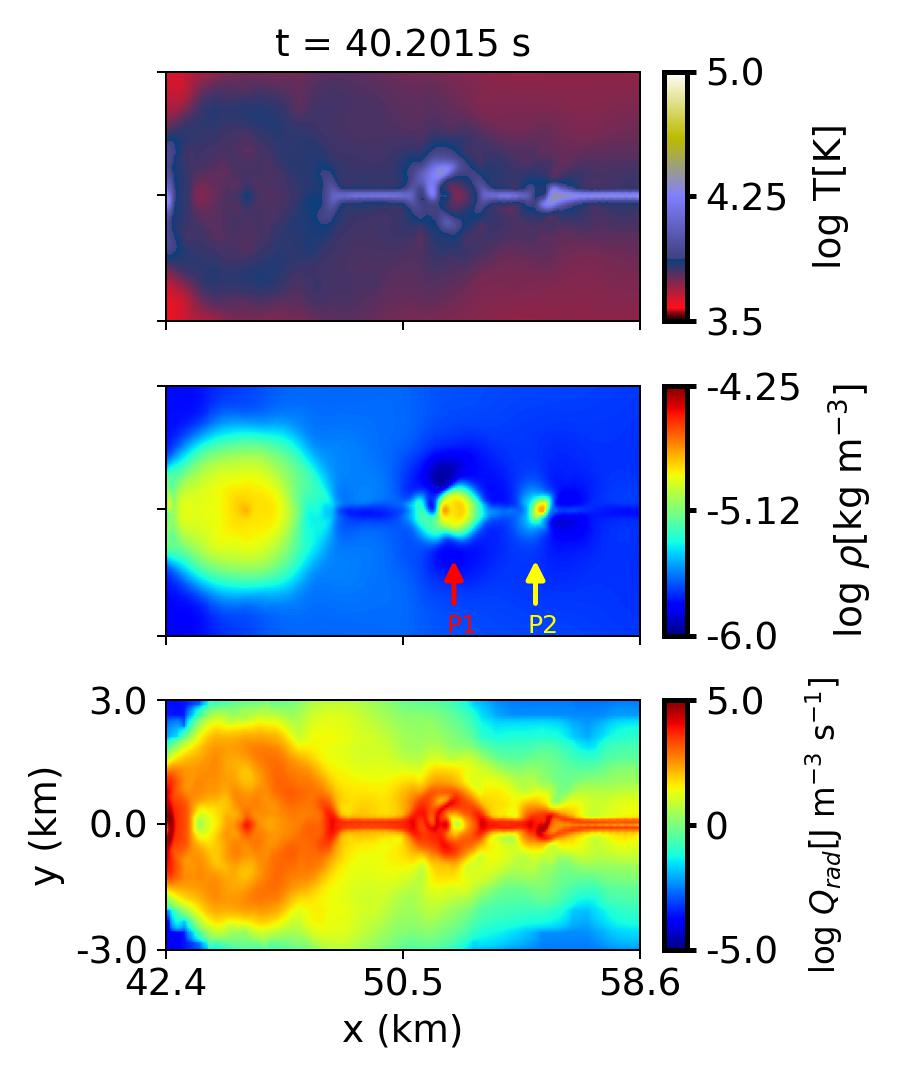}
\put(-60,165){\textcolor{white}{\textbf{(d)}}}
\end{minipage}
\begin{minipage}{0.3\textwidth}
\includegraphics[width=1.0\textwidth]{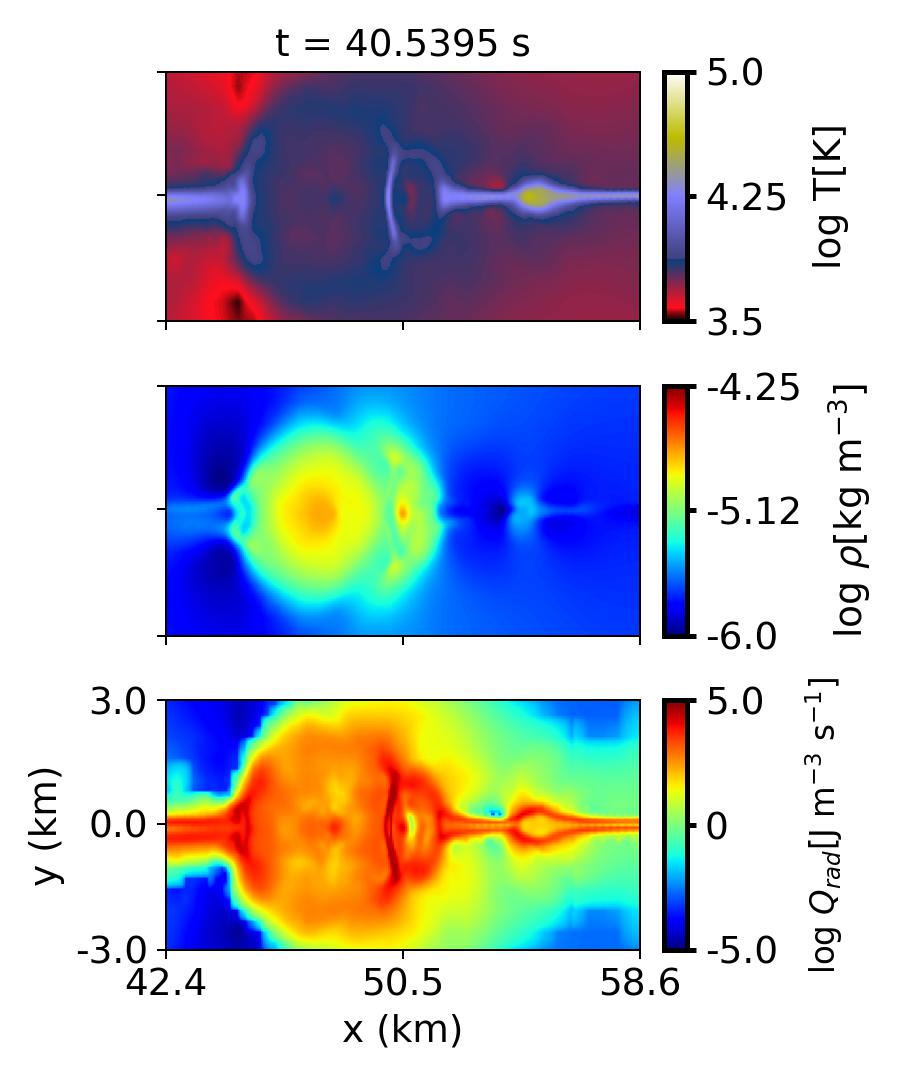}
\put(-60,165){\textcolor{white}{\textbf{(e)}}}
\end{minipage}
\begin{minipage}{0.3\textwidth}
\includegraphics[width=1.0\textwidth]{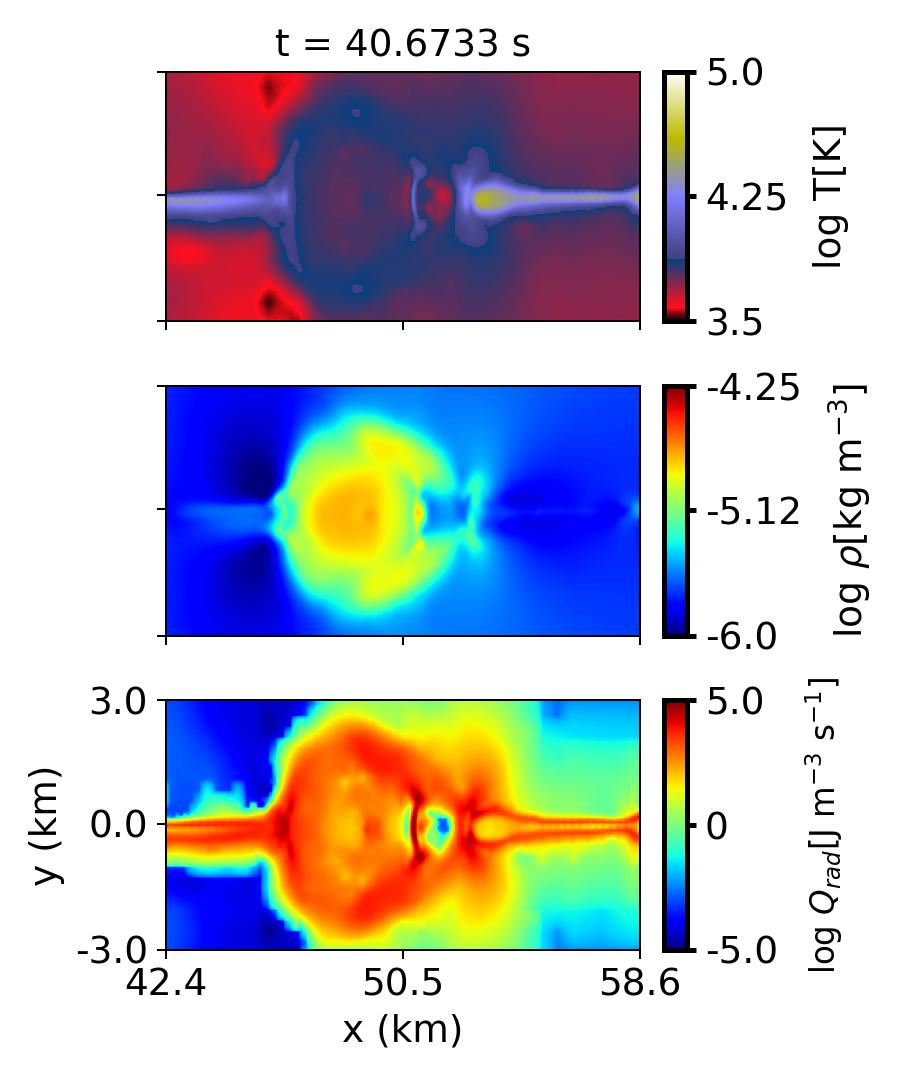}
\put(-60,165){\textcolor{white}{\textbf{(f)}}}
\end{minipage}
\caption{The evolution of the zoomed-in region within the red dotted box in Figure~\ref{fig_1}. For each time, the top, middle, and bottom panels show temperature (T), density ($\rho$), and radiative cooling ($Q_{rad}$ ), respectively. The red arrows mark the first newly formed plasmoid P1, while the yellow arrows mark the second small plasmoid P2 that appears later to the right of P1. An animation of this figure is available in the online version of the article. The animation spans the simulation interval from $t$ = 39.149 s to 40.673 s and shows the evolution of dense patches within plasmoids P1 and P2, including the later expansion-dominated cooling stage. The playback duration is $\sim$ 2 s.\\
(An animation of this figure is available.)}
\label{fig_2}
\end{figure}

The zoomed-in evolution of the newly developed plasmoid within the region enclosed by red dotted box in Figure~\ref{fig_1} is shown in Figure~\ref{fig_2}, which covers roughly from 42.4 km to 58.6 km along the $x$-axis and $-3$ km to +3 km along the $y$-axis. From top to bottom, each snapshot displays the distributions of temperature, density, and radiative cooling.
This sequence highlights the thermodynamic evolution of a newly developed plasmoid from its early stage to the production of a dense-cold clump, and eventually its interaction with a big plasmoid, that enters the zoomed-in region from the left.


At $t = 39.1497$ s, the newly developed plasmoid (hereafter referred to as P1) is located around $x = $49.218 (Figure~\ref{fig_2}(a)). P1 at this stage is relatively hot, with a maximum temperature of around $49, 000$ K. 
Its peak density reaches roughly $4.07 \times 10^{-6}$ kg m$^{-3}$ , slightly larger than the initial plasma density $\rho_0 \simeq 2.4 \times 10^{-6}$ kg m$^{-3}$. 
Even at the edges, the plasma temperature is roughly $25, 000$ K. Unlike the surrounding current sheet, the density accumulation on right face of P1 leads to a dense structure. The initial increase of density is due to the reconnection outflow from the X-point to the right of the plasmoid, which supplies plasma into this dense region. The resulting density enhancement increases the local radiative cooling and lowers the temperature. 
The decreasing of the temperature might further amplify the radiative cooling effect, resulting in additional temperature depletion. Such a runaway-like cooling process may continue until the plasma enters a lower-temperature region, where a further temperature drop is accompanied
by a decrease in radiative cooling, leading the sequence to breakdown.


\begin{figure}
\centering
\begin{minipage}{0.49\textwidth}
\includegraphics[width=1.0\textwidth]{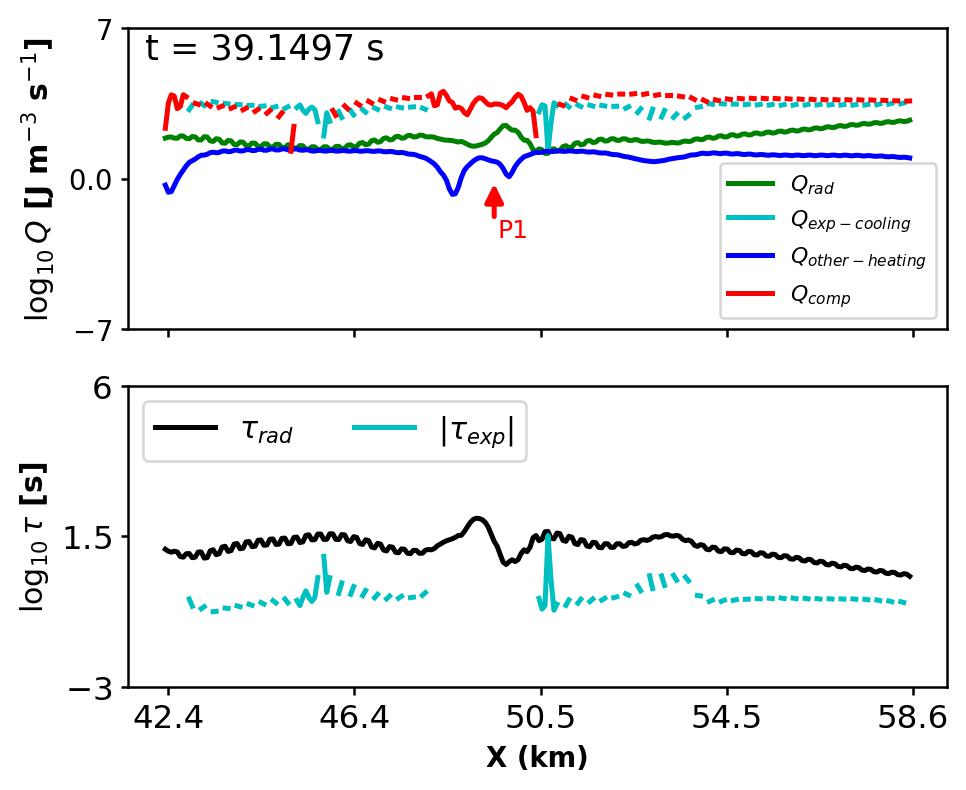}
\put(-25,188){\textbf{(a)}}
\end{minipage}
\begin{minipage}{0.49\textwidth}
\includegraphics[width=1.0\textwidth]{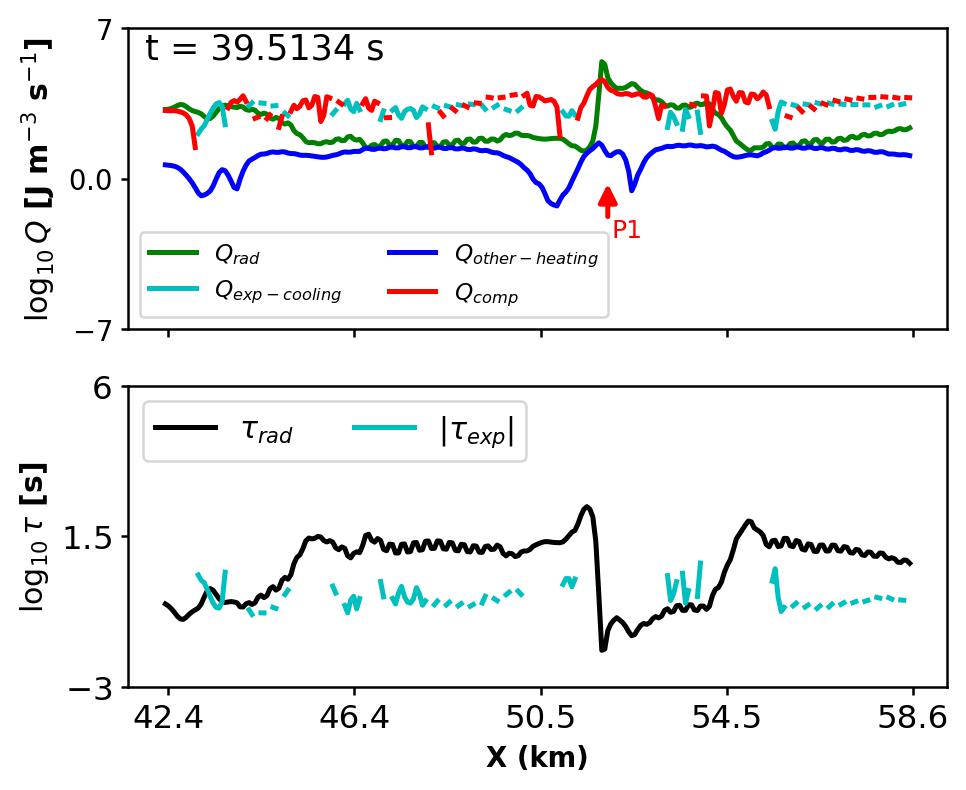}
\put(-25,188){\textbf{(b)}}
\end{minipage}
\begin{minipage}{0.49\textwidth}
\includegraphics[width=1.0\textwidth]{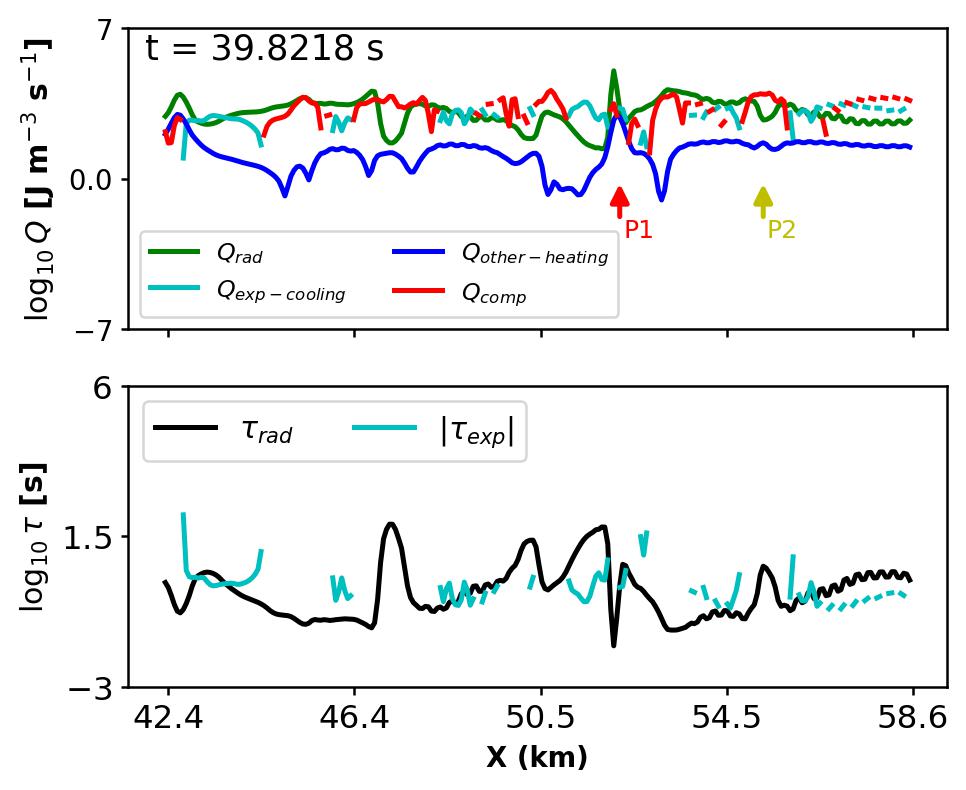}
\put(-25,188){\textbf{(c)}}
\end{minipage}
\begin{minipage}{0.49\textwidth}
\includegraphics[width=1.0\textwidth]{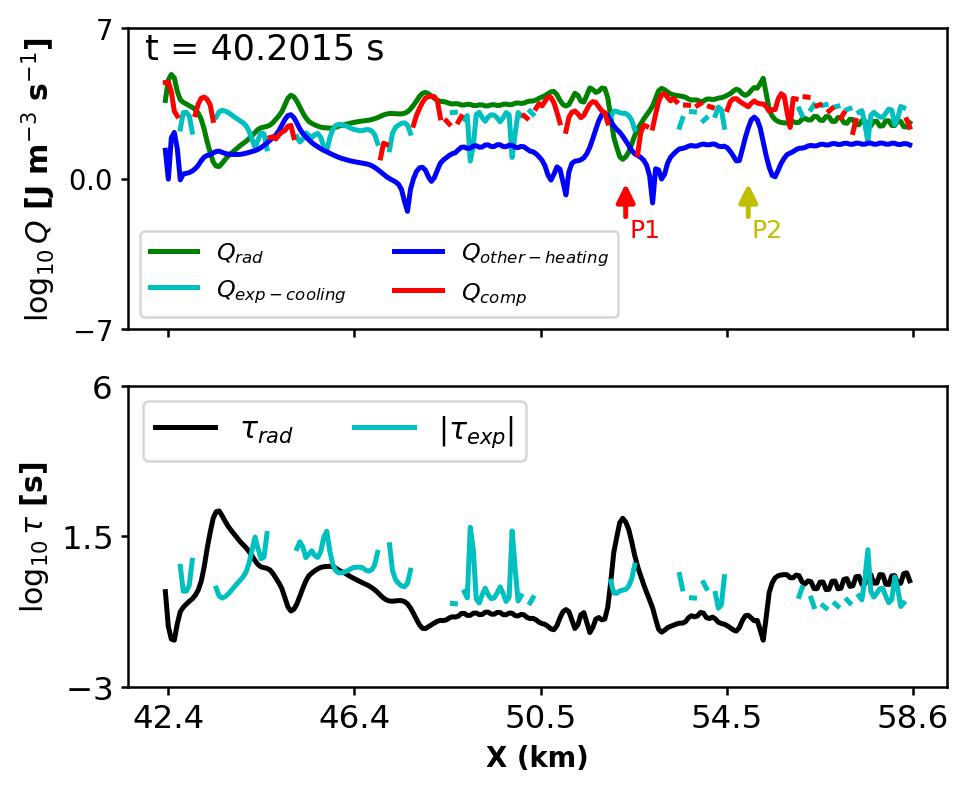}
\put(-25,188){\textbf{(d)}}
\end{minipage}
\begin{minipage}{0.49\textwidth}
\includegraphics[width=1.0\textwidth]{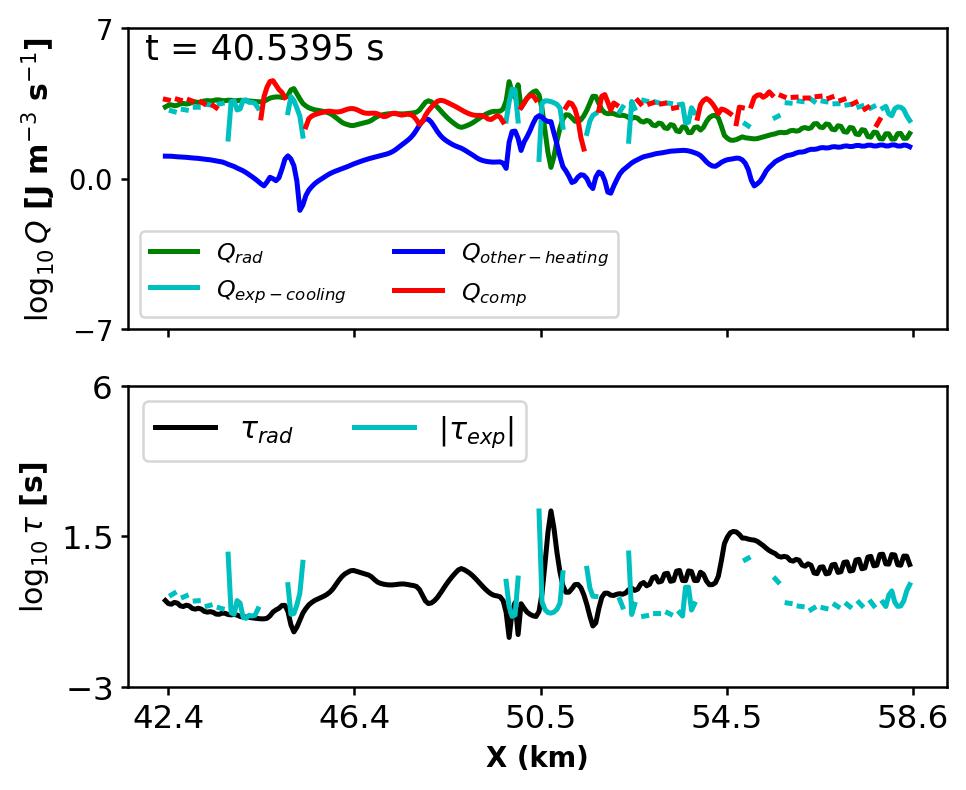}
\put(-25,188){\textbf{(e)}}
\end{minipage}
\begin{minipage}{0.49\textwidth}
\includegraphics[width=1.0\textwidth]{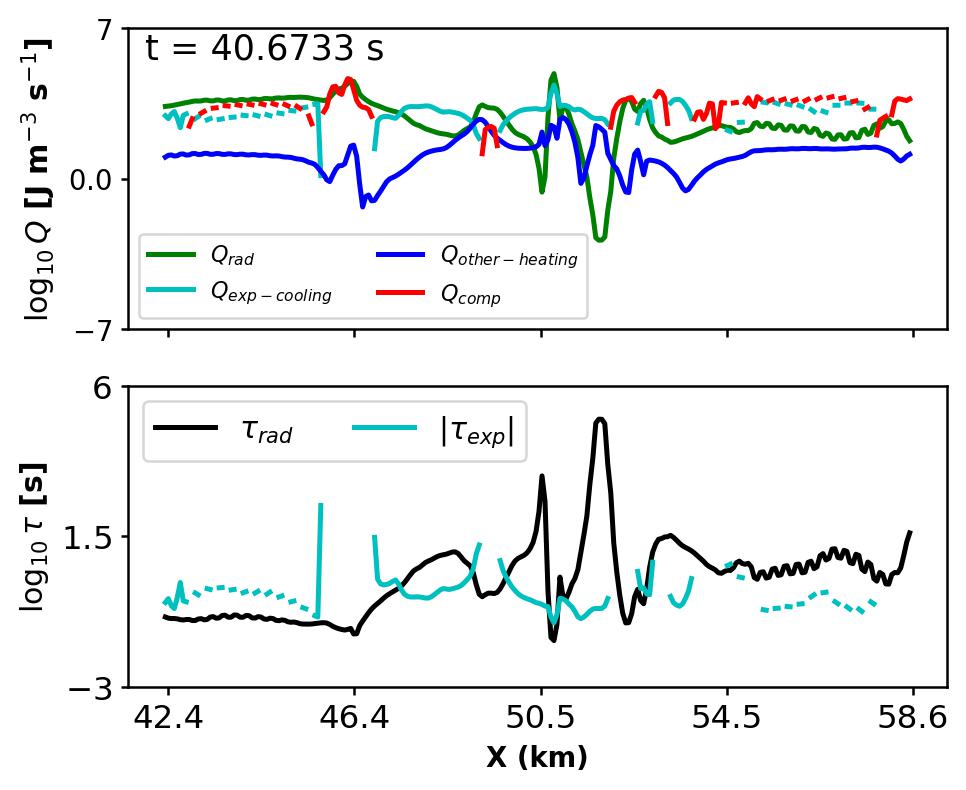}
\put(-25,188){\textbf{(f)}}
\end{minipage}
\caption{One-dimensional profiles along $y = 0$ in the zoomed region shown in Figure~\ref{fig_2}. For each time, the upper panel shows the distributions of radiative cooling (Q$_{rad}$), expansion-driven cooling (Q$_{exp-cooling}$), other heating (Q$_{other-heating}$) and
compression heating (Q$_{comp}$), while the lower panel shows the radiative cooling timescale ($\tau_{rad}$) and absolute expansion cooling
timescale ($|\tau_{exp}|$). The red and yellow arrows point to P1 and P2, respectively.}
\label{fig_3}
\end{figure}

\begin{figure}
\centering
\begin{minipage}{0.49\textwidth}
\includegraphics[width=1.0\textwidth]{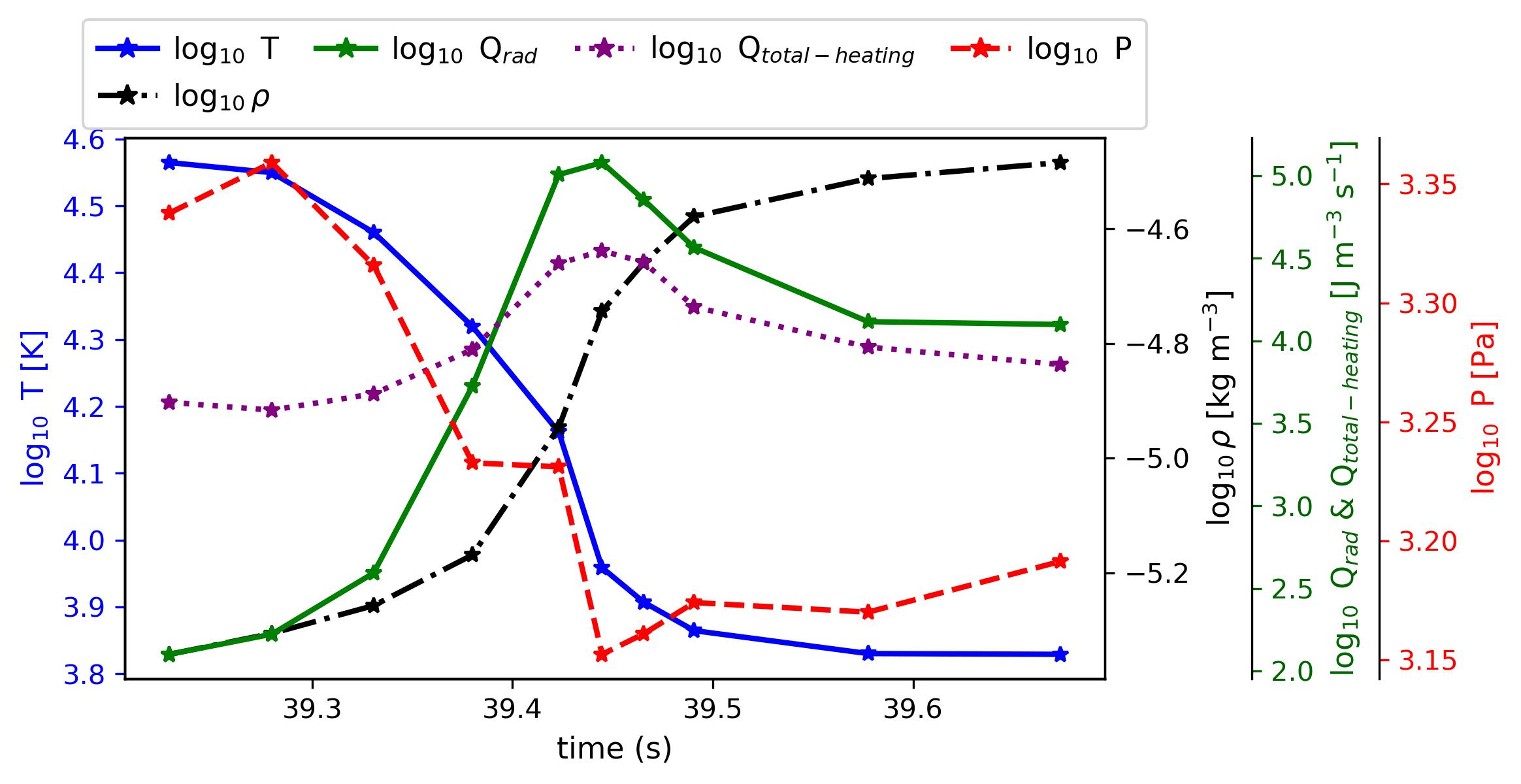}
\put(-180,22){\textcolor{black}{\textbf{(a)}}}
\end{minipage}
\begin{minipage}{0.49\textwidth}
\includegraphics[width=1.0\textwidth]{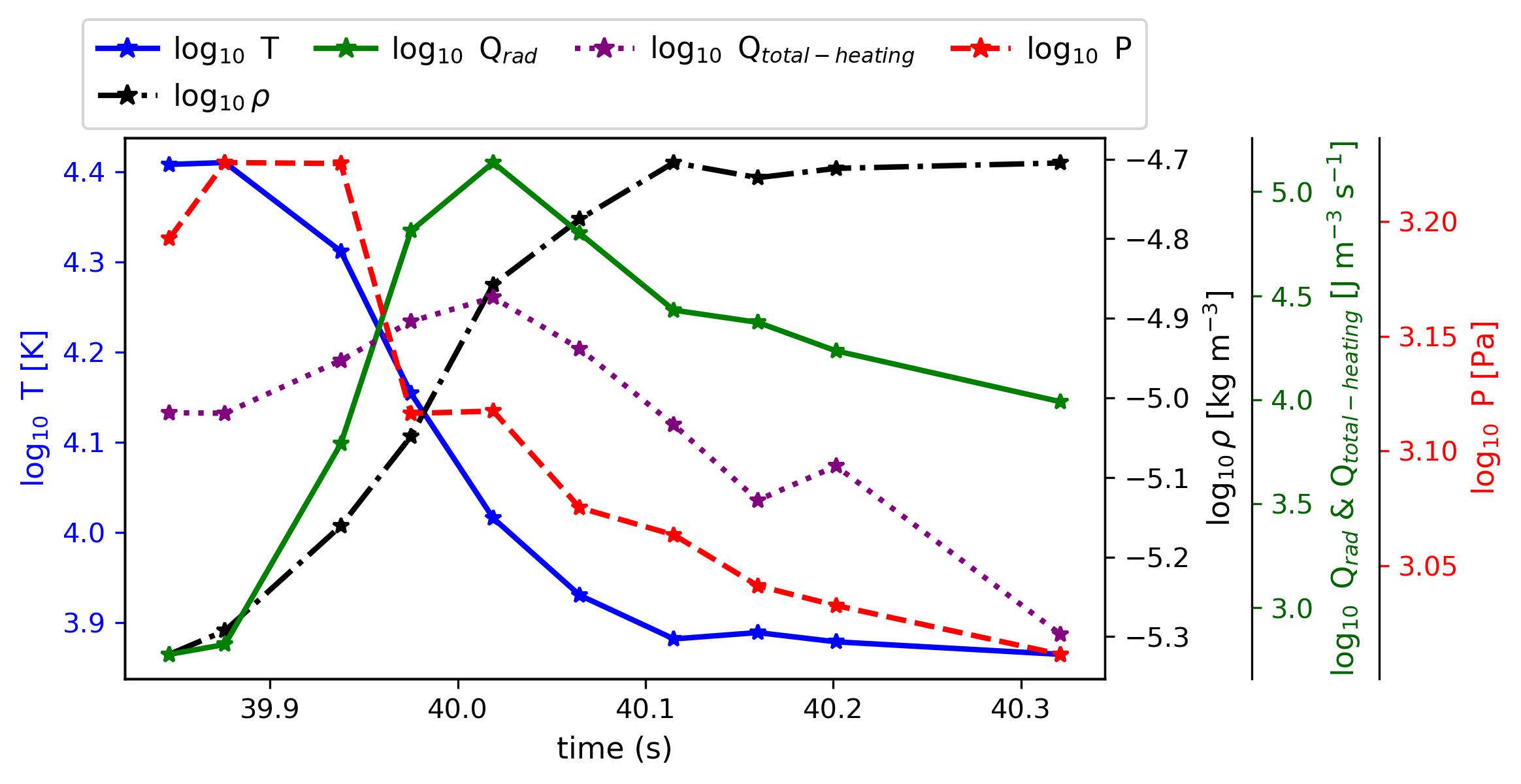}
\put(-180,22){\textcolor{black}{\textbf{(b)}}}
\end{minipage}

\caption{Time evolution of temperature, density, radiative cooling, and total heating (Q$_{comp}$ $+$ Q$_{other-heating}$) at the maximum-density location within the dense patches inside two plasmoids. Panel (a) corresponds to the results of the dense structure in P1, and panel (b) corresponds to those in P2. At each time, the maximum-density grid point is determined, and the corresponding values of T, $\rho$, Q$_{rad}$ , and Q$_{total-heating}$ are extracted at the same grid point.}
\label{fig_4}
\end{figure}

In Figure~\ref{fig_2}(b), the density of the right-edge clump reaches approximately $2.7080 \times 10^{-5}$ kg m$^{-3}$ , which is an order of magnitude larger than the initial plasma density.  
At the same spatial location, the temperature drops to around 7500 K. 
The Q$_{rad}$ panel shows that radiative cooling is significantly enhanced in this dense area. 
As time progresses, the dense-cool structure on the right edge of P1 becomes more obvious. 
The density further rises to about $3.415 \times 10^{-5}$ kg m$^{-3}$ , while the temperature drops to approximately 6100 K in the same spatial location (Figure~\ref{fig_2}(c)), which is still higher than plasma initial temperature (T$_0$ = 4421 K) but significantly lower than the plasma temperature in their surrounding. 
This lowest recorded temperature is also relatively low when compared to the plasma temperature within the newly developed plasmoid (P1), which was above $25, 000$ K in most parts of the P1. 
The coexistence of high density, low temperature, and significantly strong radiative cooling at the right edge of P1 suggests that radiative
losses become more important as the dense clump grows, supporting a transient thermal instability-like cooling process within the dense clump. 
A hot blob (plasmoid) is spotted on the left side of P1 (Figure~\ref{fig_2}(c)). 
It merges immediately with the big plasmoid to its left, making its thermodynamic behavior difficult to follow. Initially, this big plasmoid
was located slightly outside the red dotted box shown in Figure~\ref{fig_1}. It gradually moves along the x-axis to the right and eventually starts to enter the zoomed region from the left.


On the right side of P1, at around $x \simeq 55.3385$ km, another small plasmoid, P2, appears (Figure~\ref{fig_2}(c)). P2 has comparable qualitative behavior to P1: the plasmoid is initially hot, the plasma then accumulates on one side of the plasmoid, density rises locally, resulting in high Q$_{rad}$ and a reduction in temperature. At its left edge, a dense-cool plasma clump with significant Q$_{rad}$ is clearly visible in Figure~\ref{fig_2}(d).
The relatively isolated small plasmoids show a density increase accompanied by significantly strong radiative cooling and notable temperature depletion in the same region, supporting the interpretation that the dense-cool structures are most likely due to transient thermal instability-like mechanism.

The two small plasmoids (P1 and P2) initially merge and then interact with the big plasmoid (Figure~\ref{fig_2}(e), see animation also), which has moved further into the zoomed region; the plasma temperature decreases even more during this coalescence processes. The merged system evolved further, and we can see one large structure (big plasmoid) in the whole zoomed domain (Figure~\ref{fig_2}(f)). 
During this later stage, the temperature at the right edge of the big plasmoid reaches the lowest observed value of roughly 4600 K,which is slightly higher than the initial plasma temperature of 4421
K. 
At the same location, both plasma density and radiative cooling are quite low, with $\nabla \cdot V > 0$ (the contour is not shown here), indicating that this cooling is not driven by radiation losses, but by expansion associated with the merging dynamics. Thus, the whole process suggests two obvious cooling stages: first, runaway cooling phase triggered by a transient thermal instability-like mechanism in the isolated small plasmoids, where temperature decreases from several tens of thousands of kelvin to much lower values while density and Q$_{rad}$ increase; second, an additional temperature drop due to expansion during plasmoid merging.

The thermal energy density equation in the MHD model is expressed as follow:
\begin{eqnarray}
\frac{d e_{th}}{dt} = -p \nabla \cdot \mathbf{v} + \frac{1}{2 \xi} Tr(\tau^{2}_{S}) + \frac{\eta_{ei,en}}{\mu_{0}} |\nabla \times \mathbf{B}|^{2} + \frac{\eta_{AD}}{\mu_{0}^{2}} |\mathbf{B} \times (\nabla \times \mathbf{B})|^{2} + Q_{rad}. 
\label{eq:en_density}
\end{eqnarray}
where $e_{th}$ corresponds to the thermal energy density. Depending on the sign of $\nabla \cdot \mathbf{v}$, the first term on the right side represents heating/cooling induced by compression/expansion of plasma, Q$_{comp/exp} = -p\nabla \cdot \mathbf{v}$. 
The second, third, and fourth terms represent viscous heating, Joule heating (caused by the collisions of electrons with both ions and neutrals), and ambipolar heating, respectively, while the last term corresponds to radiative cooling. 
The spatial distributions of radiative cooling Q$_{rad}$, expansion cooling Q$_{exp-cooling}$, other heating Q$_{other-heating}$ (the sum of viscous heating, Joule heating, and ambipolar heating), and compression heating Q$_{comp}$ (top panels), as well as the radiative cooling timescale $\tau_{rad}$ and absolute expansion timescale $|\tau_{exp}|$ (bottom panels), along the current sheet at $y=0$ for the zoomed region shown in Figure~\ref{fig_2} are illustrated in Figure~\ref{fig_3}.
The radiative cooling and expansion cooling timescales are defined
as: $\tau_{rad} = P/[(\gamma-1)Q_{rad}]$ and $\tau_{exp} = 1/[-(\gamma-1)(\nabla \cdot \mathbf{v})]$~\citep{sen2024eruption,sen2025merging}. P1,
the first formed plasmoid, is shown by red arrows, while P2, the second tiny plasmoid that develops to the right of P1, is indicated by yellow arrows. The discontinuity in Q$_{comp}$ and Q$_{exp-cooling}$ profiles arises due to separating the compression heating and expansion-driven cooling contributions according to the local flow divergence; at a particular location, only one of these terms is present, while the other is masked.

\begin{figure}
\centering
\begin{minipage}{0.7\textwidth}
\includegraphics[width=1.0\textwidth]{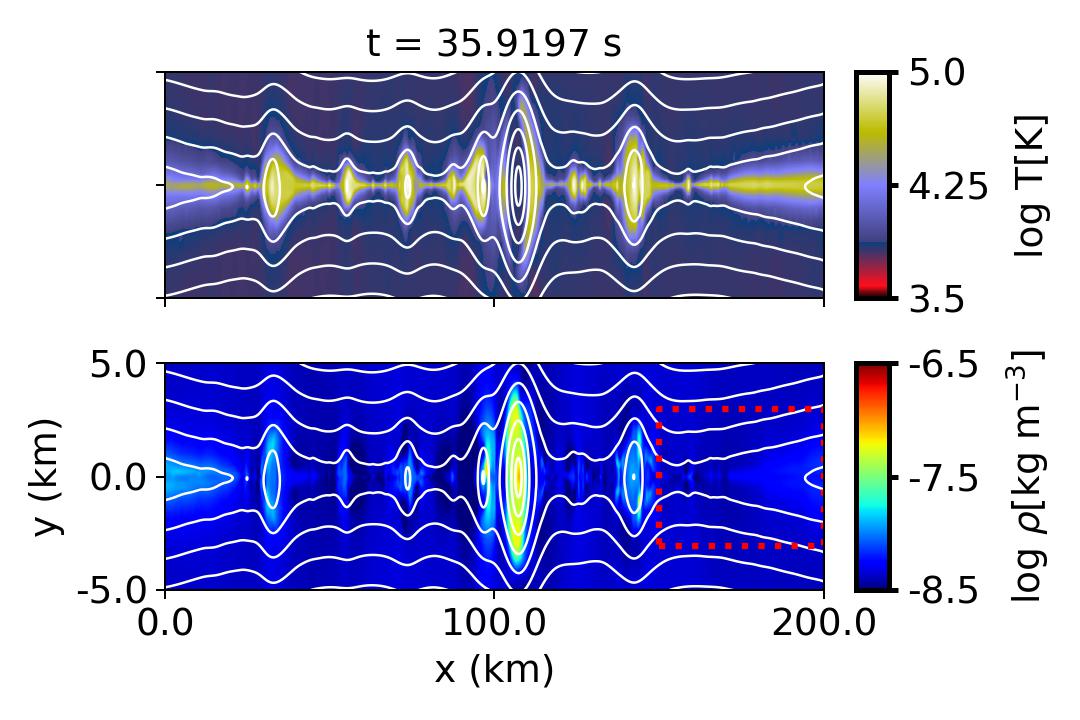}
\put(-300,205){\textcolor{white}{\textbf{(a)}}}
\end{minipage}
\begin{minipage}{0.3\textwidth}
\includegraphics[width=1.0\textwidth]{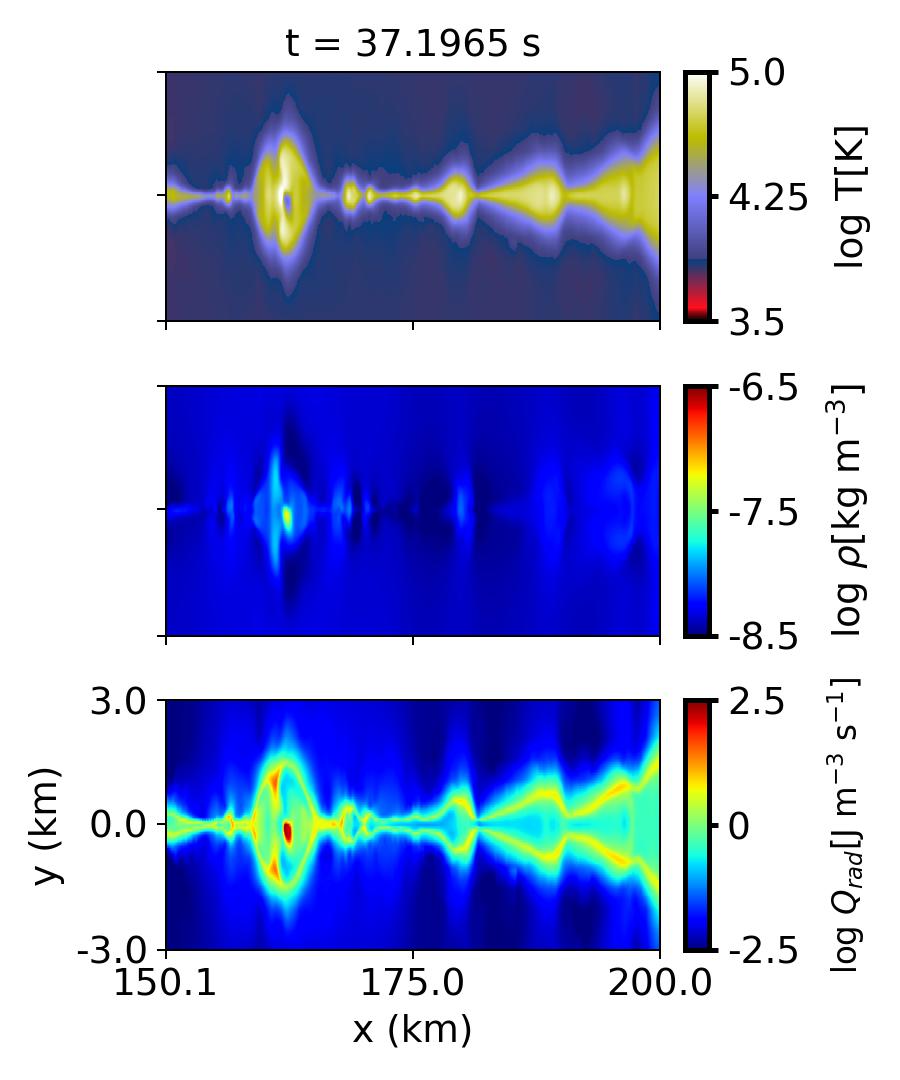}
\put(-125,165){\textcolor{white}{\textbf{(b)}}}
\end{minipage}
\begin{minipage}{0.3\textwidth}
\includegraphics[width=1.0\textwidth]{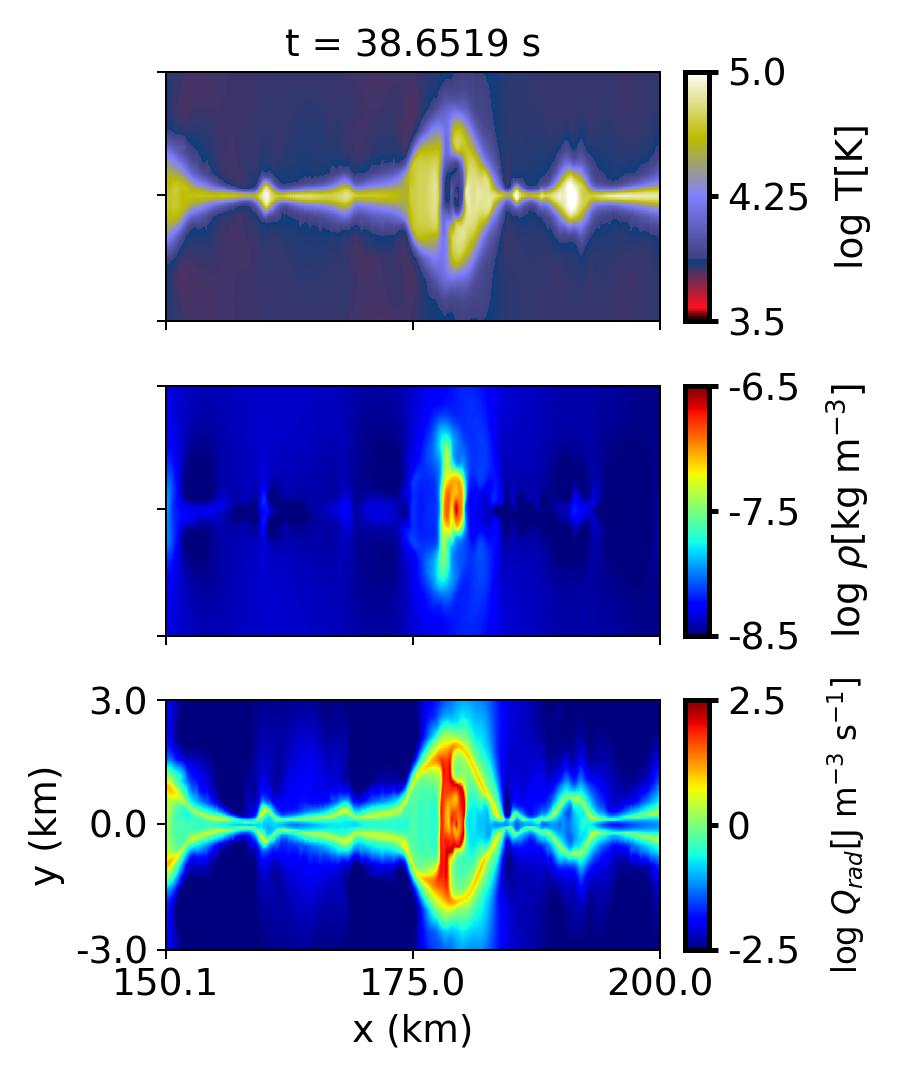}
\put(-125,165){\textcolor{white}{\textbf{(c)}}}
\end{minipage}
\begin{minipage}{0.3\textwidth}
\includegraphics[width=1.0\textwidth]{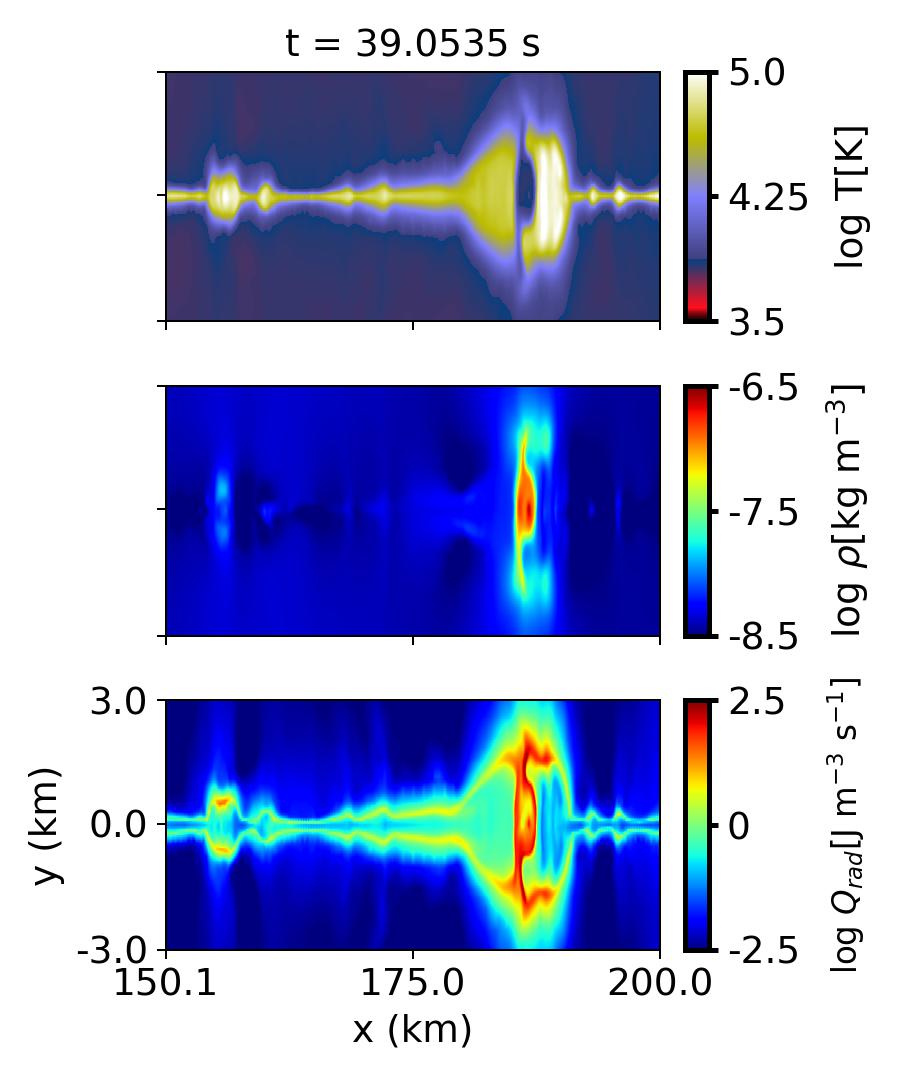}
\put(-125,165){\textcolor{white}{\textbf{(d)}}}
\end{minipage}
\caption{Representative evolution of dense structures within the plasmoid at the middle chromospheric altitude of $Z = 1400$ km with initial plasma-$\beta = 0.05$. Panel (a) shows the global distributions of temperature T in the top panel and density $\rho$ in the bottom panel; white lines indicate magnetic field lines. The red dotted box marks the selected region shown in the zoomed-in panels. Panels (b)-(d) show the zoomed-in evolution of the region of interest. At each time, the top, middle, and bottom panels show temperature T, density $\rho$, and radiative cooling Q$_{rad}$ , respectively. The zoomed-in panels show localized density enhancement, temperature decrease, and enhanced radiative cooling associated with plasmoid evolution. An animation of this figure is available in the online version of the article. The animation covers the simulation interval from $t$ = 35.484 s to 39.053 s and shows the evolution of localized dense, cool structures within the plasmoid. The playback duration is $\sim$ 1 s.\\
(An animation of this figure is available.)}
\label{fig_5}
\end{figure}

Figure~\ref{fig_3}(a) shows that Q$_{comp}$ is larger than Q$_{rad}$ at the position marked by the red arrow. This is consistent with the temperature contour in Figure~\ref{fig_2}(a), where P1 is hot, with temperatures higher than $25, 000$ K in most of P1. 
At such high temperatures, the plasma is almost fully ionized, thus Joule heating from electron-neutral collisions and ambipolar diffusion heating is minimal, resulting in a relatively small Q$_{other-heating}$ at this stage. The lower panel of Figure~\ref{fig_3}(a) demonstrates that $\tau_{rad}$ is large at this position, indicating that radiative cooling is not yet effective enough to influence the thermal evolution of the plasmoid.
Later, the situation changes substantially: Q$_{rad}$ grows significantly and becomes stronger than Q$_{comp}$ (or Q$_{total-heating}$ = Q$_{other-heating}$ + Q$_{comp}$) at the P1 position, and $\tau_{rad}$ also decreases drastically (Figures~\ref{fig_3}(b) and~\ref{fig_3}(c)). 
This coincides with the cool-dense patch at P1 (Figures~\ref{fig_2}(b) and~\ref{fig_2}(c)). 
The spatial coincidence of strong radiative cooling and significantly small radiative cooling timescale suggests that radiative cooling influences the local thermodynamics and leads to a transient thermal-instability-like cooling within the plasmoid. 
We use temperature-dependent ionizations of hydrogen and helium, therefore as the plasma cools, ionization decreases while number of neutrals increases. Hence, large number of neutrals enhances Q$_{other-heating}$ in the cool-dense clumps (Figures~\ref{fig_3}(c) and~\ref{fig_3}(d)), restricting the cooling of the plasma to an even lower temperature.
At $t = 40.2015$ s (Figure~\ref{fig_3}(d)), a dip in the radiative cooling profile (green line) is observed at P1. Q$_{exp-cooling}$ has larger amplitude than Q$_{rad}$ , and $|\tau_{exp}| < \tau_{rad}$ at the same spatial location. These characteristics demonstrate clear local contribution from expansion cooling. The corresponding contour plot in Figure~\ref{fig_2}(d) reveals that P1 stays relatively isolated at this time period. Therefore the expansion cooling is likely connected to the increasing size of P1 rather than to plasmoid coalescence dynamics. The yellow arrow points to P2, which develops to the right of P1 at
$t \sim 39.8218$ s. At this stage, the amplitude of compression heating (red line) is larger than radiative cooling (green line), suggesting that P2 is currently in its initial development phase. As P2 evolves, Q$_{rad}$ gets stronger and exceeds Q$_{comp}$ (Figure~\ref{fig_3}(d)), consistent with the cold, dense, and highly radiative region that develops along its left face in Figure~\ref{fig_2}(d).

\begin{figure}
\centering
\begin{minipage}{0.7\textwidth}
\includegraphics[width=1.0\textwidth]{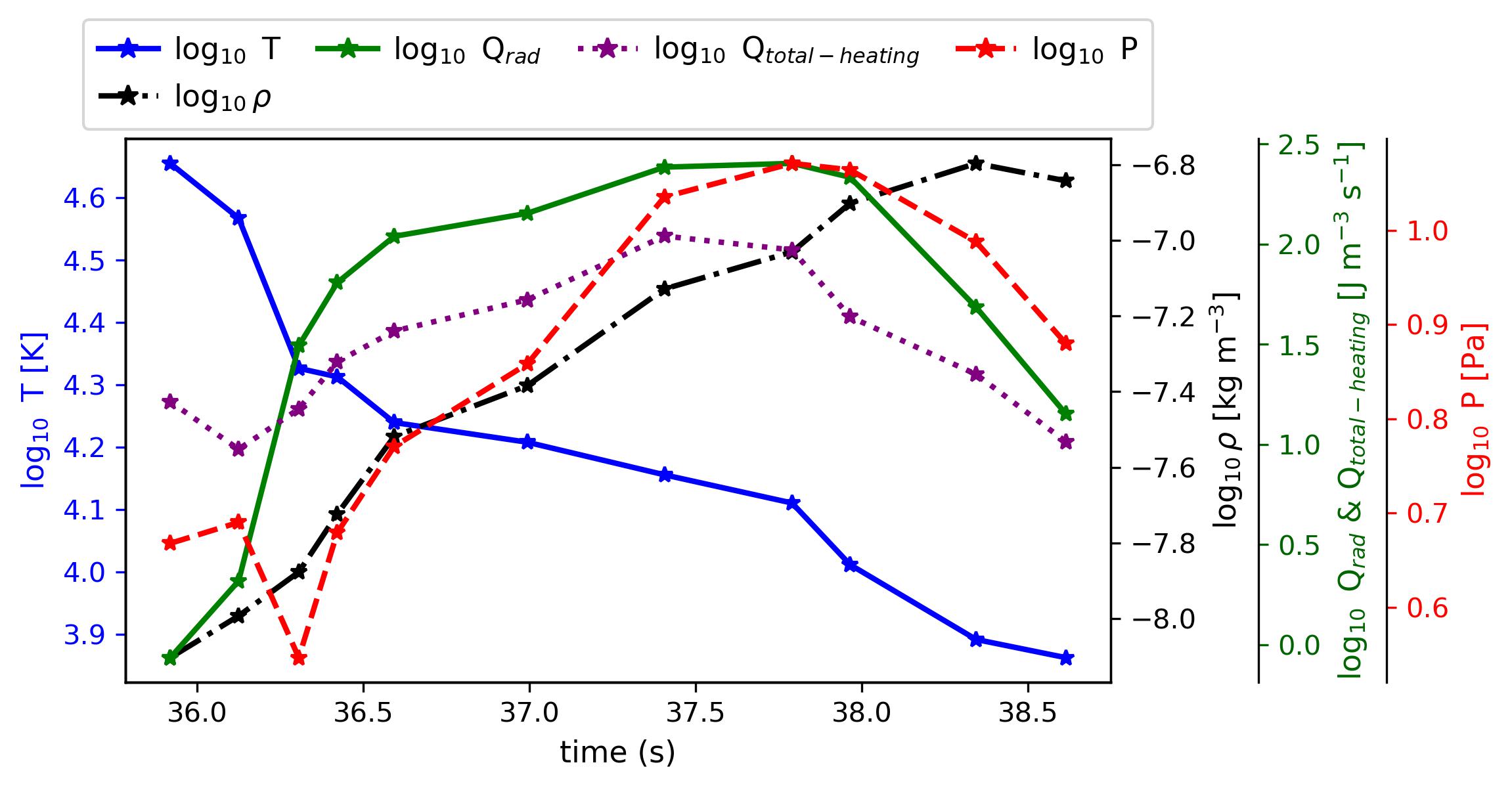}
\end{minipage}
\caption{Temporal evolution of T, $\rho$, P, Q$_{rad}$ , and Q$_{total-heating}$ at the maximum-density grid point within the dense structure at $Z = 1400$ km with $\beta_0 = 0.05$.}
\label{fig_6}
\end{figure}

At $t = 40.5395$ s and $t = 40.6733$ s, noticeable dips in Q$_{rad}$ are observed near $x \sim 50.7$ km and $x \sim 51.7$ km, respectively. 
These spatial locations approximately correlate to the right face of the big plasmoid seen in Figures~\ref{fig_2}(e) and~\ref{fig_2}(f). 
The lower panels of Figures~\ref{fig_3}(e) and~\ref{fig_3}(f) show that the expansion cooling timescale is much shorter than the radiative cooling timescale at these dips. 
This suggests that radiative cooling is no longer efficient enough to contribute to plasma cooling in these areas. 
In contrast, the shorter expansion timescale indicates that expansion cooling becomes more relevant when the plasmoids interact. 
In these expansion-driven cooling regions, Q$_{other-heating}$ again contributes to local plasma heating, counteracting the cooling caused by the expansion and preventing the plasma temperature from dropping to significantly low values. 
These analyses are consistent with the contour sequence in Figure~\ref{fig_2}, where the temperature drops further during the later merging stage, reaching its lowest values after the onset of expansion-dominated evolution.

\begin{figure}
\centering
\begin{minipage}{0.7\textwidth}
\includegraphics[width=1.0\textwidth]{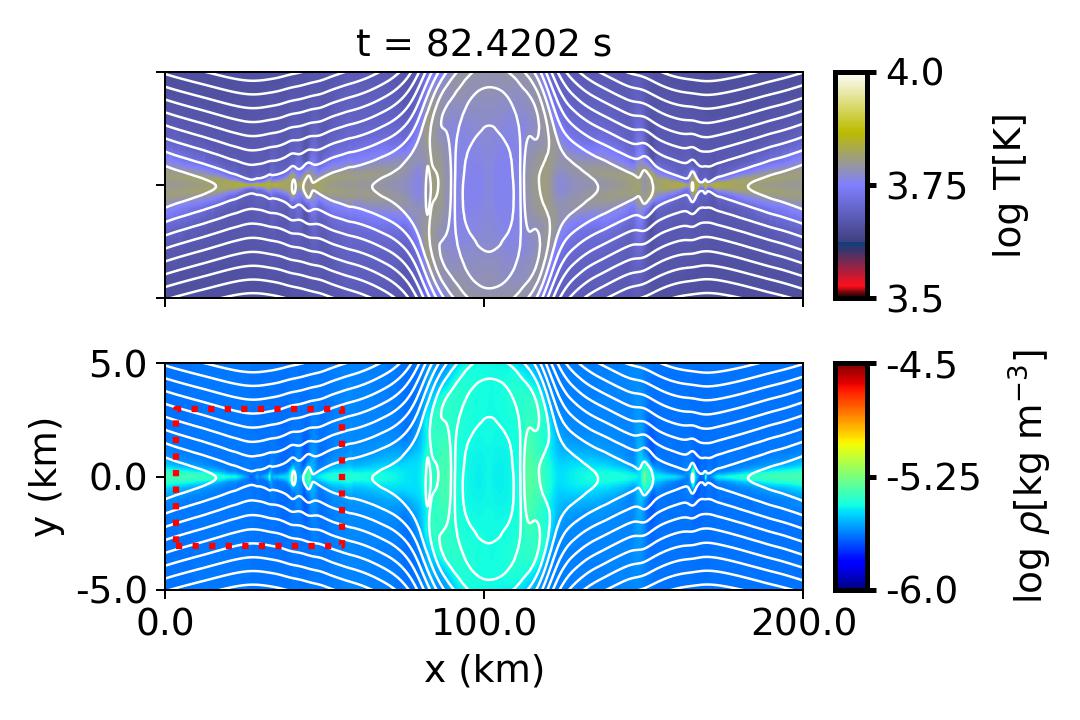}
\put(-300,205){\textcolor{white}{\textbf{(a)}}}
\end{minipage}
\begin{minipage}{0.3\textwidth}
\includegraphics[width=1.0\textwidth]{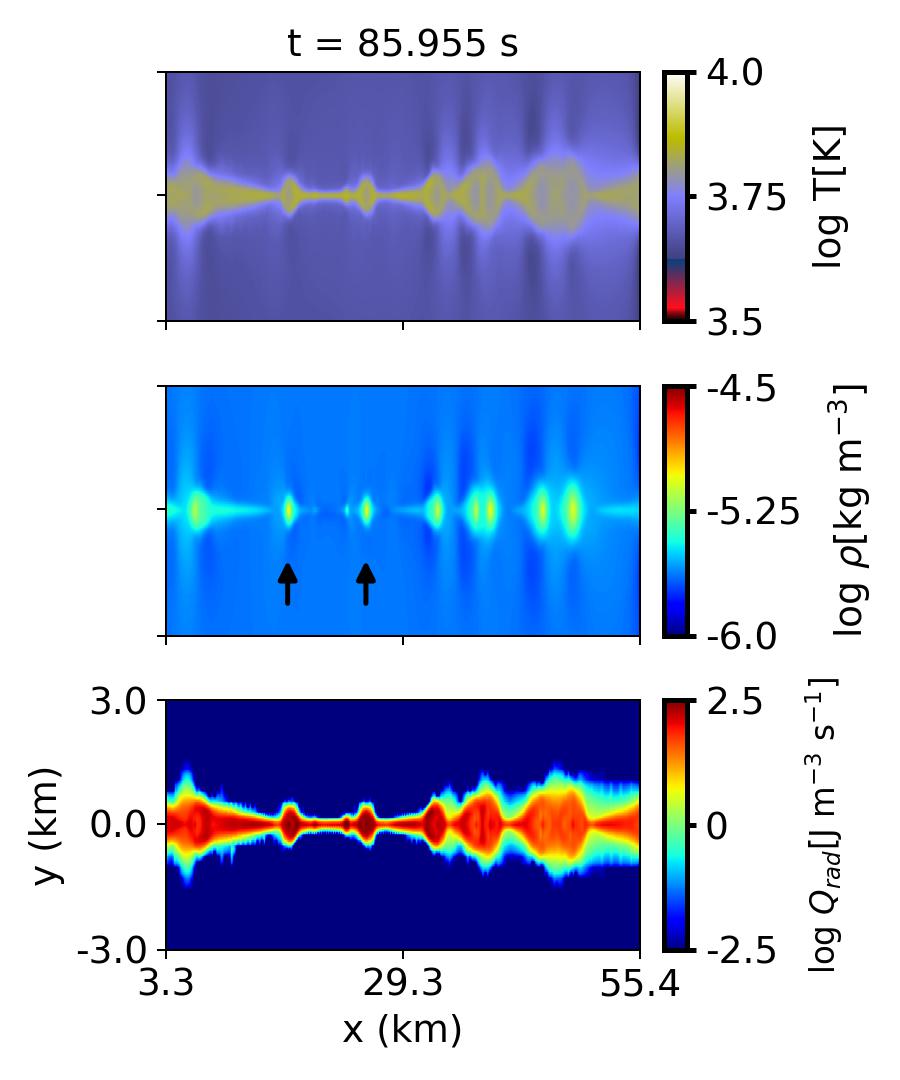}
\put(-125,165){\textcolor{white}{\textbf{(b)}}}
\end{minipage}
\begin{minipage}{0.3\textwidth}
\includegraphics[width=1.0\textwidth]{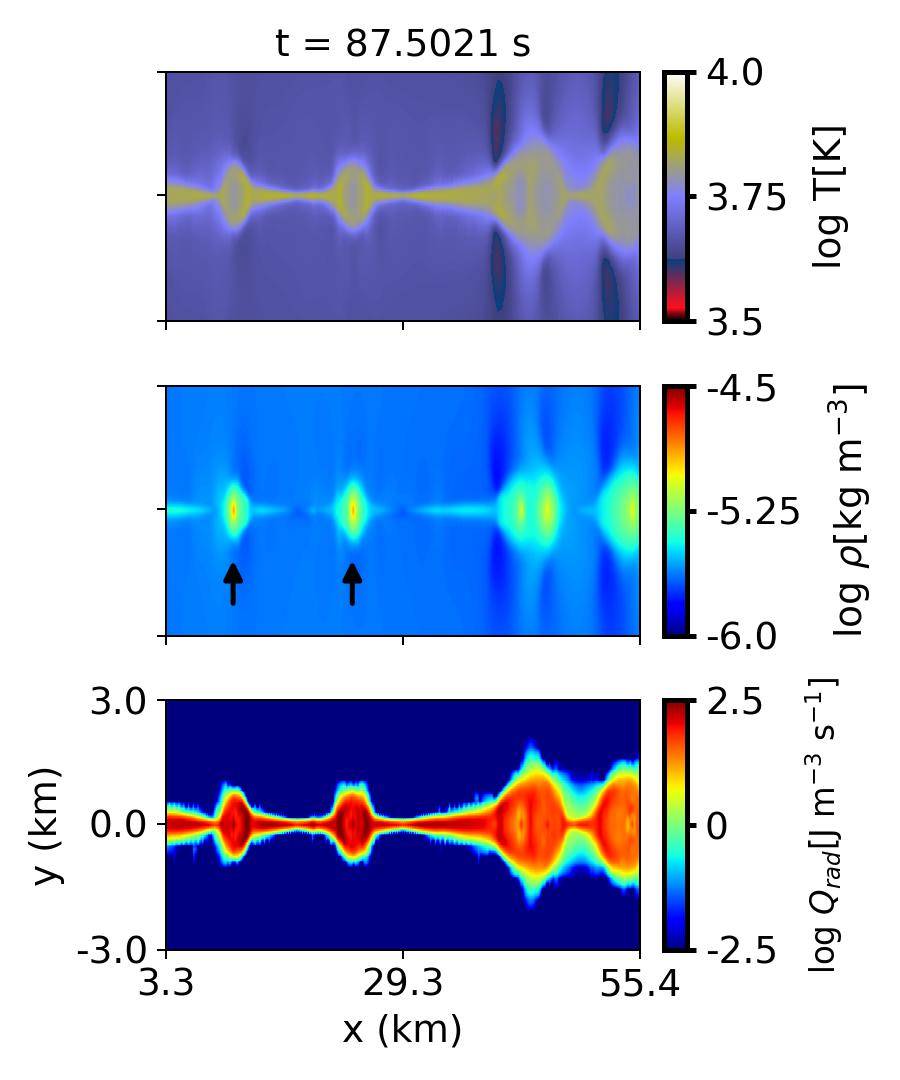}
\put(-125,165){\textcolor{white}{\textbf{(c)}}}
\end{minipage}
\begin{minipage}{0.3\textwidth}
\includegraphics[width=1.0\textwidth]{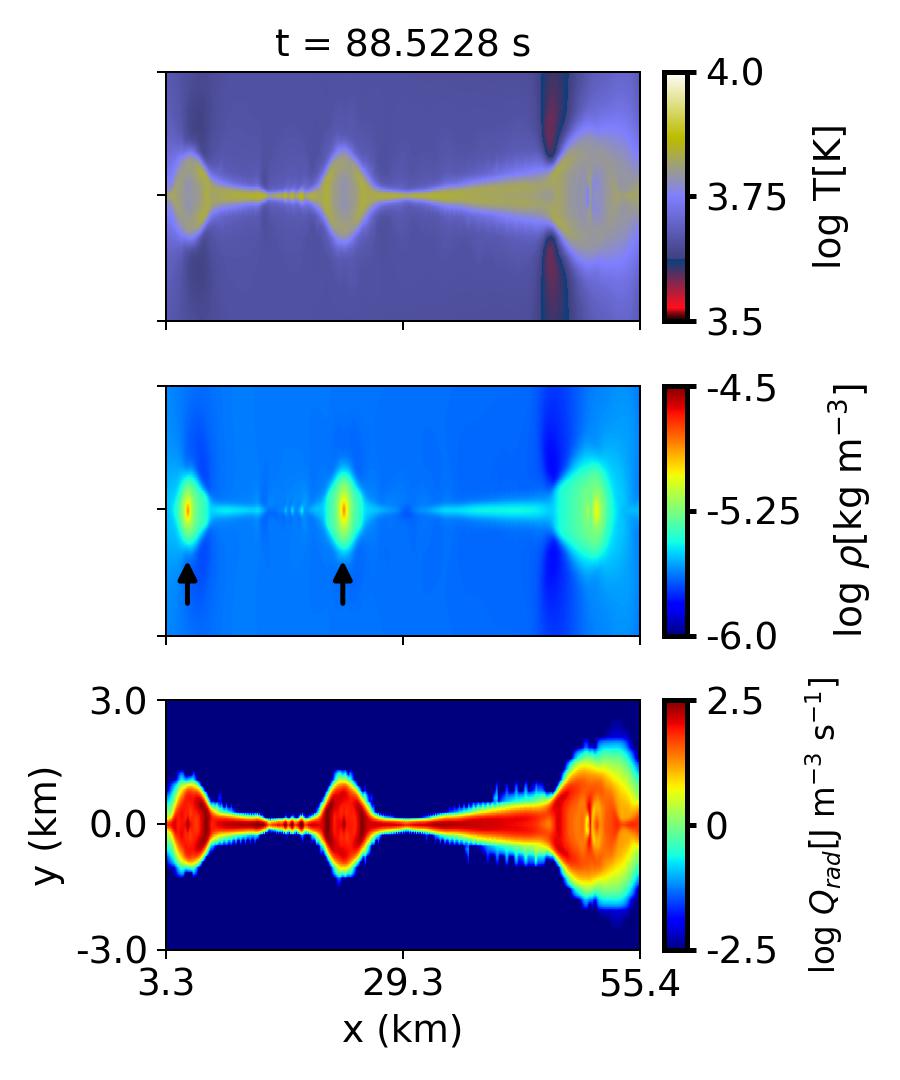}
\put(-125,165){\textcolor{white}{\textbf{(d)}}}
\end{minipage}
\caption{Evolution of plasmoid-associated density enhancement and cooling in the high-$\beta_0$ simulation of 0.5 at $Z = 600$ km. Panel (a) shows the global distributions of temperature (top panel) and density(bottom panel), with white lines indicating magnetic field lines. The red dotted box marks the selected region shown in the zoomed-in panels. Panels (b)-(d) show the zoomed-in evolution of this region at selected times. For each time, the top, middle, and bottom panels show temperature, density, and radiative cooling, respectively. The black arrows mark dense patches inside the plasmoid. Although localized density enhancement and cooling are present, the dense structures remain relatively cool from the beginning and do not show the catastrophic temperature decrease found in the low-$\beta_0 = 0.05$ case.}
\label{fig_7}
\end{figure}

Figure~\ref{fig_3} verifies the two-stage cooling illustration inferred from Figure~\ref{fig_2}. 
During the early development of P1 and P2, the compression heating or total heating (Q$_{comp}$ $+$ Q$_{other-heating}$) is larger than radiative cooling. 
However, as these plasmoids evolve, radiative cooling exceeds the heating terms while $\tau_{rad}$ becomes shorter, indicating that radiative cooling becomes dominant as dense clumps form within the plasmoids. 
Then, the local dips in the radiative cooling profiles occur when isolated plasmoids grow in size and/or interact with surrounding plasmoids. 
At these locations, the expansion cooling timescale becomes shorter than the radiative cooling timescale, making expansion cooling more effective than radiative cooling during this later stage.

\begin{figure}
\centering
\begin{minipage}{0.49\textwidth}
\includegraphics[width=1.0\textwidth]{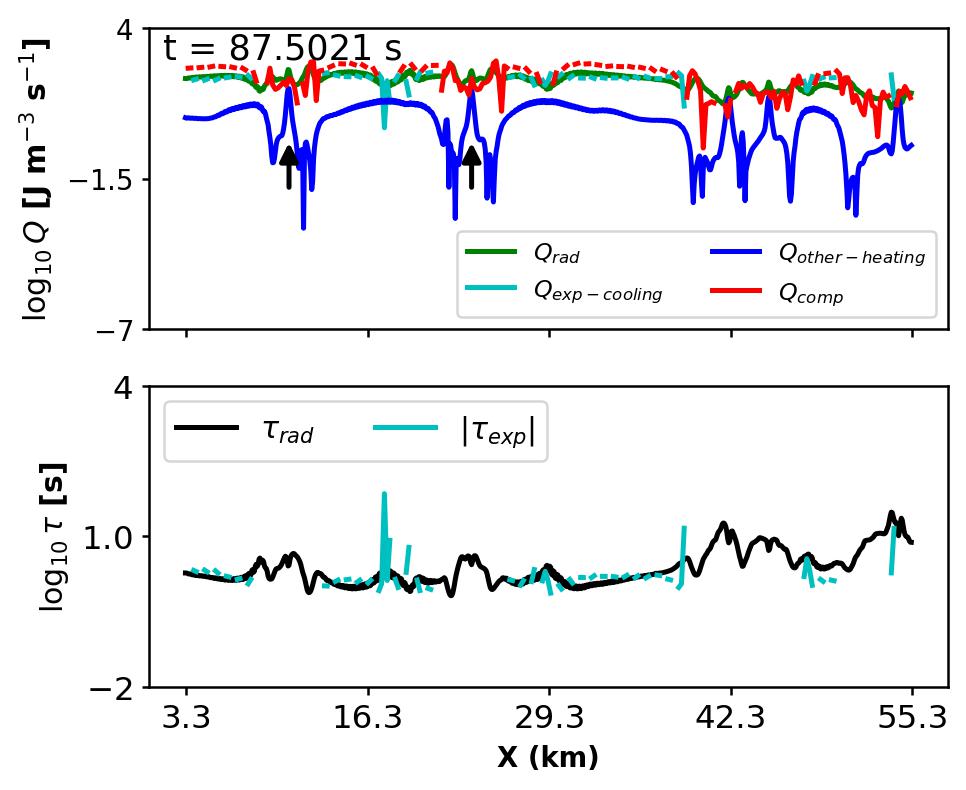}
\put(-25,188){\textbf{(a)}}
\end{minipage}
\begin{minipage}{0.49\textwidth}
\includegraphics[width=1.0\textwidth]{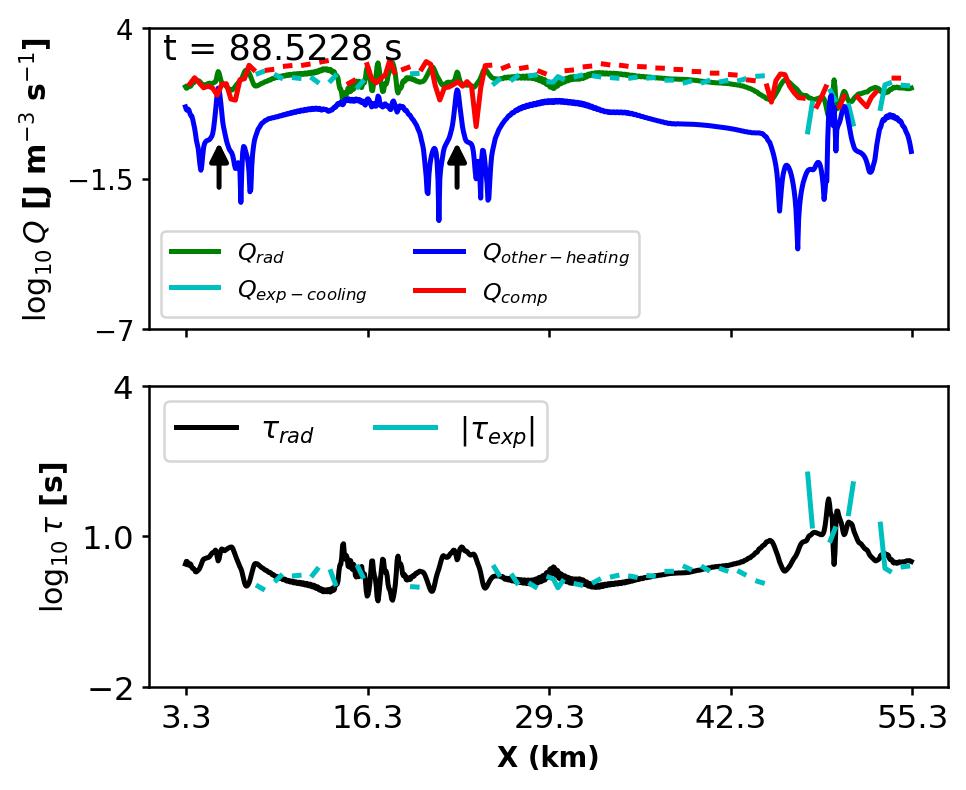}
\put(-25,188){\textbf{(b)}}
\end{minipage}
\begin{minipage}{0.7\textwidth}
\includegraphics[width=1.0\textwidth]{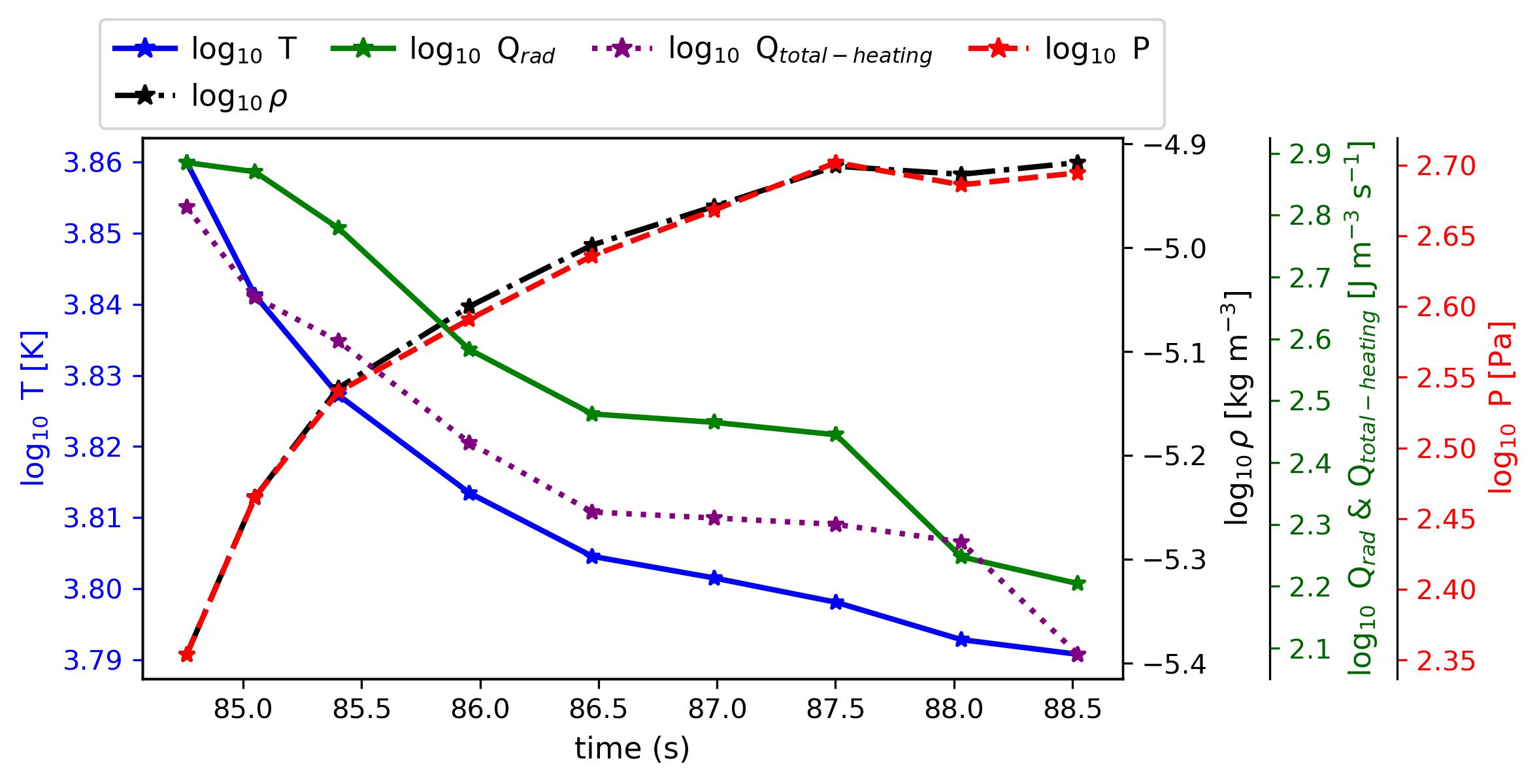}
\put(-310,30){\textbf{(c)}}
\end{minipage}
\caption{Diagnostic profiles and temporal evolution of the dense patch in the $\beta_0 = 0.5$ simulation case at $Z = 600$ km. Panels (a)-(b) show one-dimensional profiles along $y = 0$ at the selected times shown in Figure~\ref{fig_7} (c, d). For each time, the upper panel shows radiative cooling Q$_{rad}$ , expansion cooling Q$_{exp-cooling}$ , other heating Q$_{other-heating}$ and compressional heating Q$_{comp}$ , while the lower panel shows the radiative cooling timescale $\tau_{rad}$ and the absolute expansion cooling timescale $|\tau_{exp}|$. The black arrows mark the dense-patch locations identified in Figure~\ref{fig_7}. Panel (c) shows the temporal evolution of T, $\rho$, P, Q$_{rad}$ , and Q$_{total-heating}$ at the maximum-density grid point within the dense core of the plasmoid. In contrast to the low-$\beta_0 = 0.05$ case, T and Q$_{rad}$ decrease together, indicating gradual cooling rather than an approximately isobaric runaway-cooling phase.}
\label{fig_8}
\end{figure}

\begin{figure}
\centering
\begin{minipage}{0.6\textwidth}
\includegraphics[width=1.0\textwidth]{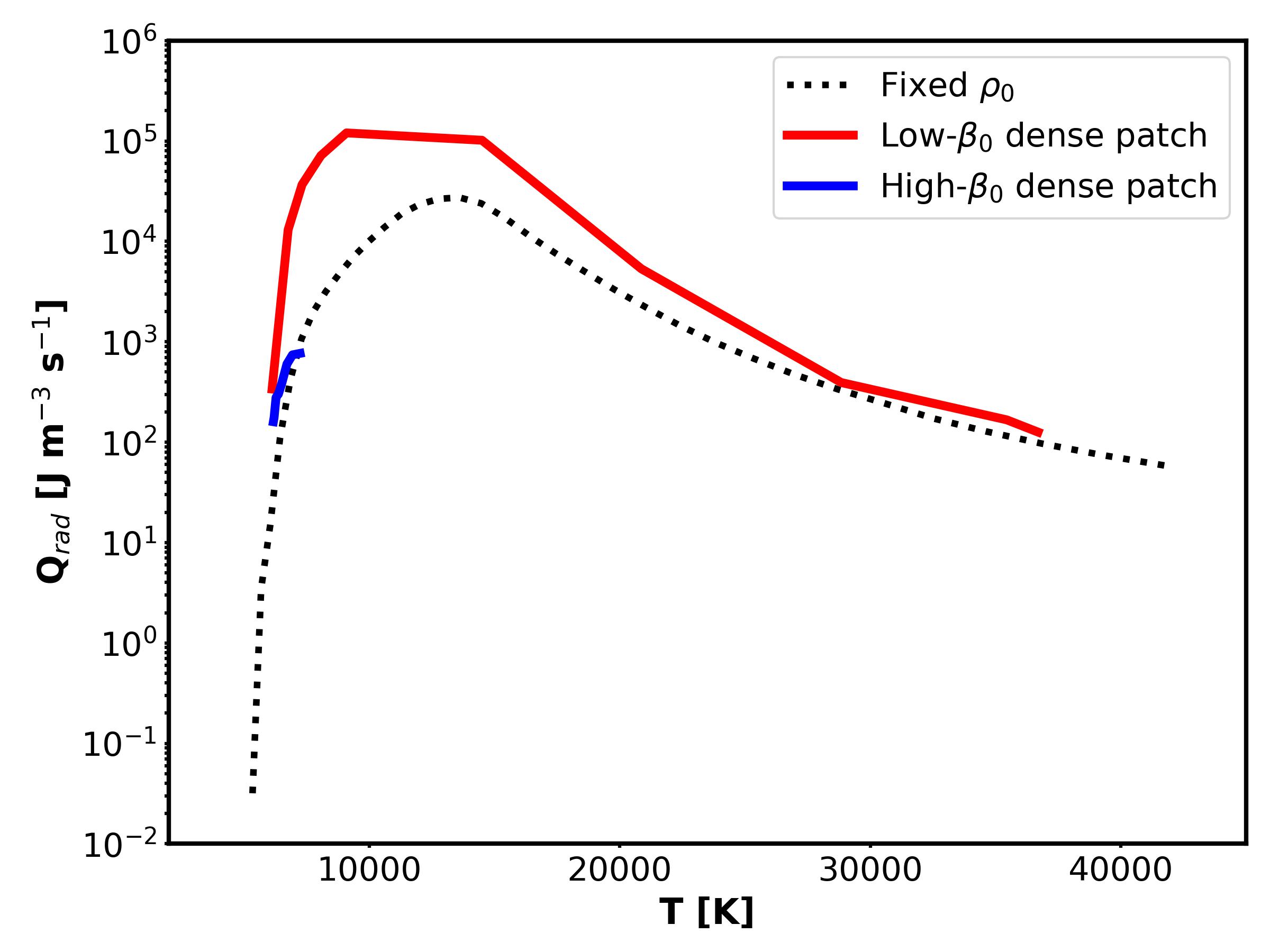}
\end{minipage}
\caption{Comparison of radiative cooling behavior as a function of temperature for the $Z = 600$ km cases. The dotted black curve shows Q$_{rad}$ calculated at the fixed initial density $\rho_0 = 2.40 \times 10^{-6}$ kg m$^{-3}$. The data for the red curve are extracted from the dense structure in P1 in the $\beta_0 = 0.05$ simulation, whereas the blue curve corresponds to the evolution of the dense core in the $\beta_0 = 0.5$ simulation. The low-$\beta_0$ curve passes through the temperature range where Q$_{rad}$ increases as T decreases, favoring runaway radiative cooling. In contrast, the high-$\beta_0$ curve remains mainly below $\sim 8000$ K, where further cooling is not accompanied by increasing Q$_{rad}$.}
\label{fig_9}
\end{figure}

Figure~\ref{fig_4} shows the time evolution of various plasma parameters at the density peak point inside the dense patch identified in Figure~\ref{fig_2}. For each plasmoid, we first identified the region of the plasmoid where the dense patch develops.
At each time, the maximum-density grid-point within this specified region was determined, and the temperature, density, pressure, radiative cooling rate, and total heating (Q$_{total-heating}$ $=$ Q$_{comp}$ $+$ Q$_{other-heating}$ ) were extracted from that position. 
Before $t = 40.2015$ s, P1 has not interacted with other nearby plasmoids (Figure~\ref{fig_2}(d)), yet we observed signatures of expansion-cooling (Figure~\ref{fig_3}(d)), likely linked to the increase in its size.  
Therefore, to examine the radiative-cooling-dominated evolution, we follow the time evolution only up to the stage before expansion-cooling becomes significant.
Figure~\ref{fig_4}(a) demonstrates
the thermodynamic evolution of P1, showing that at the beginning of the time series, the temperature is high, whereas the density, total heating, and radiative cooling rate are low. Later, the plasma density sharply increases and the temperature decreases. Before $t = 39.38$ s, Q$_{total-heating}$ is stronger than Q$_{rad}$ , but Q$_{rad}$ rises sharply and exceeds Q$_{total-heating}$ as time passes. Such an explosive increase in radiative cooling is accompanied by a sharp drop in temperature and a sharp rise in density over a short period, thus supporting the interpretation of a transient thermal instability-like process that leads to cool-dense structure.


Figure~\ref{fig_4}(b) demonstrates that P2 has a similar thermodynamic evolution as P1. Figure~\ref{fig_4} supports the concept
established from Figures~\ref{fig_2} and~\ref{fig_3}. In both plasmoids (P1 and P2), density increases while temperature falls, and Q$_{rad}$
grows considerably during a short period. As shown in Figure~\ref{fig_4}, the plasma pressure only slightly changes during this time period, and the negative trend of radiative cooling with temperature indicates $(\partial Q_{rad}/\partial T)_{P} < 0$, consistent with the expected signature of an isobaric thermal instability-like process. The good agreement among contour plots, one-dimensional profiles, and time series plots shows that the cool-dense patches inside the plasmoids are closely linked to condensations driven by a transient thermal instability-like mechanism during the nonlinear evolution of the
reconnecting current sheet.

For initial plasma-$\beta$ of 0.05, similar behavior is observed at a middle chromospheric height of $Z = 1400$ km. At this height, the background plasma density and temperature are $\rho_0 = 4.51 \times 10^{-9}$ kg m$^{-3}$ and T$_0$ = 6610 K, respectively. The corresponding initial magnetic field strength for $\beta_0 = 0.05$ is $b_0 = 3.2 \times 10^{-3}$ T. 
In Figure~\ref{fig_5}, the top panels are the representations of the global view of the plasmoid-dominated reconnecting current sheet, while the zoomed-in panels focus on localized dense structures associated with the plasmoids. 
The coexistence of condensed structures and strong radiative cooling at the same locations is qualitatively comparable to what we observed and analyzed in greater detail for the $Z = 600$ km case. 
Figure~\ref{fig_6} shows that the thermodynamic evolution of the dense-cool structure is comparable to that of the dense-cool structures in P1 and P2. 
Similar dense-cool asymmetric structures associated with transient thermal-instability-like processes are also observed at $Z = 1000$ km with $\beta_0 = 0.05$ simulation (results not shown). 
Thus, the presence of cool-dense structures at different chromospheric altitudes suggests that the condensations triggered by transient thermal instability-like phenomena may take place under a variety of chromospheric conditions rather than being limited to a single location.

In order to determine whether the cool-dense structures linked with transient thermal instability-like process occur under weaker magnetic field conditions, we conducted a simulation at $Z = 600$ km with a higher initial plasma-$\beta$ of 0.5 ($b_0 = 1.9 \times 10^{-2}$ T). 
Figure~\ref{fig_7} displays the global and zoomed-in evolution of this case. In Figure~\ref{fig_7}(a), the global temperature and density distributions show that, similar to the low-$\beta_0$ case, the current sheet has multiple plasmoids. 
The red dotted box highlights the domain selected for a detailed investigation. The zoomed-in panels in Figures~\ref{fig_7}(b)-\ref{fig_7}(d) present the temperature, density, and radiative cooling distributions within this region of interest at particular times. 
Here the magnetic reconnection process only leads to a small temperature increase in the current sheet region, and the maximum temperature reaches a value below 8000 K in this case with $\beta_0 = 0.5$. 
As plasmoids grow larger, their cores become significantly denser, with densities occasionally reaching about an order of magnitude above the initial plasma density, while the central temperature drops to a lower value. 
However, the temperature decrease in the plasmoid center is much milder compared with the lower-$\beta$ case.

Figures~\ref{fig_8}(a) and~\ref{fig_8}(b) present one-dimensional cuts along $y = 0$ for the region enclosed by the red dotted box in the time frames corresponding to Figures~\ref{fig_7}(c) and~\ref{fig_7}(d). In contrast to simulations with low-$\beta_0$, in dense cores marked by black arrows, Q$_{rad}$ shows no significant enhancement and remains comparable to the heating terms. Furthermore, the timescales of radiative cooling remain relatively long, and thus radiative losses are not significant.


Figure~\ref{fig_8}(c) represents the time evolution of the dense core marked by the second arrow on the left in Figure~\ref{fig_7}. 
Like in the low-$\beta_0$ cases, a maximum-density grid point is determined at each time point, and the corresponding density, temperature, pressure, total heating, and radiative losses are extracted. The time evolution shows that both temperature and Q$_{rad}$ decrease concurrently in this high-$\beta_0$ case, and $(\partial Q_{rad} /\partial T)_{P} > 0$. 
Such characteristics are not consistent with the thermal instability-like phenomena identified in low-$\beta_0$ simulations. Although dense-cool cores develop inside the plasmoids in the high-$\beta_0$ simulation, the density increase and temperature drop are much more gradual. 
More importantly, the radiative cooling Q$_{rad}$ always decreases with time during this period, and the negative trend of radiative cooling with temperature as in the low-$\beta$ case is not observed, implying that a transient thermal-instability-like process does not happen here. We have examined the evolution of all plasmoids in this simulation and
found no evidence to support the existence of thermal instability in this high-$\beta$ simulation.

The above analyses showed that the formation of cool-dense asymmetric structures within the plasmoid in the low-$\beta_0$ simulations is associated with a transient thermal instability-like process, whereas the high-$\beta_0$ case displays no runaway-like cooling phases. 
To further understand the difference between the two cases and the criteria for the onset of thermal instability in a chromospheric reconnection process, we examine the variation of Q$_{rad}$ with temperature in Figure~\ref{fig_9}. 
This gives insight into why radiative cooling increases as temperature decreases in the low-$\beta_0$ case, while no such anti-correlation is seen in the high-$\beta_0$ case. 
The black dotted curve presents the variation of radiative cooling (Eq.~\ref{eq:rad}) with temperature using a fixed initial plasma density ($\rho_0 = 2.40 \times 10^{-6}$ kg m$^{-3}$ ) at $Z = 600$ km. 
The red and blue curves in Figure~\ref{fig_9} are also calculated based on Equation~\ref{eq:rad}. 
The red curve is derived from the values (temperature, density, ionization degrees) at the grid point with the maximum density inside plasmoid P1 in the case with $\beta_0 = 0.05$. As the temperature inside P1 decreases from $38, 000$ K to $6100$ K, Q$_{rad}$ initially increases by three orders of magnitude and then sharply decreases once the temperature drops below $8000$ K. 
Such a large temperature drop over a wide range leads to a runaway-like cooling process and the transient thermal-instability-like condensation. 
The short blue curve is derived from the values (temperature, density, ionization degrees) at the grid point with the maximum density inside the dense plasmoid as shown in Figure~\ref{fig_7} and Figure~\ref{fig_8} in the case with $\beta_0 = 0.5$. 
Inside this dense plasmoid, the temperature ranges from $6100$ K to $8000$ K, with a maximum of only about $8000$ K. As the temperature drops below $8000$ K, Q$_{rad}$ also decreases. 
The maximum Q$_{rad}$ in the case with $\beta_0 = 0.5$ is about two orders of magnitude smaller than that in the case with $\beta_0 = 0.05$. Therefore, the runaway-like cooling process and thermal instability do not happen in this higher-$\beta$ case.

\section{Summary and Discussion}
\label{sec-IV}
Using a series of resistive-MHD simulations that includes the non-adiabatic effect of chromospheric radiative energy loss, temperature-dependent ionizations, and viscosity, we analyzed the formation and evolution of cool-dense structures within the plasmoids during magnetic reconnection in the partially ionized chromospheric plasma. The core objective of this study was to explore whether the development of localized cool and dense structures under different initial plasma conditions ($\rho_0$ , T$_0$ ) is associated with a transient thermal-instability-like process. The primary results and conclusions are summarized below.
\begin{enumerate}
    \item In a low-$\beta$ chromospheric environment, where local magnetic reconnection heats the plasmoids to a temperature above $20, 000$ K, the growing plasmoids will usually undergo two cooling phases. During the first cooling phase, the localized accumulation of plasma and decreasing temperature lead to a runaway-like strong radiative cooling process and transient thermal-instability-like condensation. Later, the increase in size of a plasmoid and the coalescence dynamics result in the further expansion cooling phase. Such cooling phases have been identified in all lower-$\beta$ simulations from the low chromosphere to the upper chromosphere.
    \item In a high-$\beta$ chromospheric environment, when magnetic reconnection can only heat the plasma to temperatures below $10, 000$ K, thermal instability does not develop. The decreasing temperature causes a reduction in radiative cooling when the temperature is below $8000$ K, and the negative trend of radiative cooling with temperature is not observed. The plasma density is also increased inside the plasmoid, but the radiative cooling has never entered into the explosive increasing stage.
\end{enumerate}

Recent MHD studies have shown that plasmoids or flux ropes can trap plasma and develop localized condensations in the coronal current sheet~\citep{sen2022thermally,sen20233d,de2025coupled}. 
Extending the concept to partially ionized chromospheric plasma, where partial ionization, strong radiative losses, and plasma-$\beta$ effects become important, our findings show that the cool-dense structures associated with transient thermal-instability-like process emerge within the plasmoids which are initially heated to a temperature above 20, 000 K in the lower plasma-$\beta$ environment.

Such results indicate that the cool-dense structures can possibly appear in the local high temperature
solar activities such as UV bursts, which then explain why both the Si IV emissions and H$\alpha$ wing emissions can be detected simultaneously in some of these UV bursts~\citep{peter2014hot,tian2016iris,chen2019flame}. 
However, we should point out that such transient thermal-instability-like processes can not happen in the lower temperature reconnection events such as Ellerman Bombs. This process might also be relevant to other partially ionized astrophysical environments, such as the interstellar medium, cometary plasmas, protoplanetary disks, and others, where magnetic fields and radiative losses could influence the triggering of thermal instability-like occurrence, which leads to the formation of cool-dense structures embedded in the hot and low density regions.

The present 2.5D simulations cannot fully capture the additional plasma dynamics that may arise in a three-dimensional configuration. In 3D, magnetic flux ropes can form within the reconnecting current sheet, and cool, dense condensations may develop within these structures~\citep{sen20233d,de2025coupled}. 
The recent 3D radiative MHD simulation~\citep{cheng2025three} also confirmed the formation of small flux ropes in the chromospheric magnetic reconnection process, and the coexistence of cool dense plasma and lower density hot plasma in the newly formed flux ropes and reconnection region is clearly shown. 
In a three-dimensional configuration, the extended structure of the flux ropes may allow additional plasma redistribution along the magnetic field. Dense material may drain along the flux rope, whereas plasma supplied from magnetically connected regions may increase the available mass and potentially sustain the condensation. Three-dimensional simulations have shown that cool, dense structures can undergo substantial field-guided motion and mass redistribution within complex magnetic configurations~\citep{xia2016formation,kohutova2020self}. 
Therefore, the condensation process identified here must also occur within three-dimensional flux ropes, but the morphology, density evolution, cooling behavior, and lifetime of the condensations may differ from those obtained from the present 2.5D simulations.
Furthermore, the gravitational stratification, current-sheet inclination, and a larger spatial extent can also modify the density distribution, plasma drainage, reconnection dynamics, and consequently the onset and lifetime of the condensations. 
Thus, our present results should not be interpreted as a quantitative prediction for arbitrary chromospheric current sheets. Simulations including gravity, atmospheric stratification, and more realistic current-sheet geometries will be necessary to determine how the condensation mechanism operates in such configurations in the future works.

The spatial and temporal resolution of the plasmoids in our simulations and the cool-dense condensations formed within them is beyond the capability of current observational instruments.  
Therefore, neither the detailed plasmoid structure nor the embedded condensation is likely to be directly identified as a resolved feature in present observations.
High-resolution multi-wavelength observations from ground and space-based facilities, such as DKIST~\citep{rimmele2020daniel} and the proposed SCOPE mission~\citep{lin2025solar}, might identify signatures of dense-cool structures similar to those observed in our numerical experiments. 
However, plasmoid-mediated reconnection can produce characteristic spectroscopic and imaging spectropolarimetric signatures associated with complex velocity and density structures, while the rapid density enhancement and temperature decrease of the embedded condensations may further modify the emission from the unresolved reconnection region and produce detectable intensity or spectral signatures in temperature-sensitive diagnostics~\citep{leenaarts2025high}.
Previous studies have shown that unresolved plasmoid-mediated reconnection can produce characteristic spectroscopic signatures, including complex Si IV profiles with strong line cores and broad wings~\citep{innes2015iris}, as well as broad, non-Gaussian, and triangular-shaped Si IV and Ca II profiles~\citep{rouppe2017intermittent}. 
The recent advanced imaging spectropolarimetric data taken in the He I 1083 nm line also provide the evidence of small plasmoid-like structures in the chromosphere~\citep{leenaarts2025high}.
More recently,~\citet{cheng2024magnetic} synthesized Si IV and H\(\alpha\) profiles from simulations of plasmoid-mediated reconnection in the lower solar atmosphere, further illustrating how such diagnostics
can connect complex small-scale reconnection structures with observable signatures. 
Future work should therefore generate synthetic multi-wavelength and spectroscopic diagnostics from our simulations and account for the spatial, temporal, and spectral resolution of the relevant instruments, allowing both the signatures of plasmoid-mediated reconnection and the additional effects of the embedded cool-dense condensations to be tested observationally.
\begin{acknowledgments}
This research is supported by the National Key R\&D Program of China Nos. 2022YFF0503800 (2022YFF0503804) and 2022YFF0503003 (2022YFF0503000); the NSFC grants 12373060, 11933009 and 12573062; the Strategic Priority Research Program of the Chinese Academy of Sciences with grant No. XDB0560000; the Basic Research of Yunnan Province in China with grant 202401AS070044; the Yunling Talent Project for the Youth; the Yunling Scholar Project of the Yunnan Province and the Yunnan Province ScientistWorkshop of Solar Physics; Yunnan Key Laboratory of Solar Physics and Space Science under the number 202205AG070009. SS acknowledges support by the Research Council of Finland through the Centre of Excellence project, SpaceResilience (Grant\#374096). The work has been carried out at the National Supercomputer Center in Tianjin, and the calculations are performed on the Tianhe new generation supercomputer. The data analysis has been performed at the Computational Solar Physics Laboratory of Yunnan Observatories
\end{acknowledgments}

\bibliography{sample701}{}
\bibliographystyle{aasjournalv7}



\end{document}